\documentclass[preprint,12pt,amsmath,amssymb,superscriptaddress,floatfix]{revtex4-2}

\usepackage{graphicx}
\graphicspath{{./}{./figs/}}
\usepackage{amsmath,amssymb,amsthm}
\usepackage{bm}
\usepackage{enumitem}
\usepackage{xcolor}
\usepackage{hyperref}
\usepackage{booktabs}
\usepackage{multirow}
\usepackage{float}
\usepackage{subcaption}
\usepackage{microtype}
\definecolor{phStore}{RGB}{31,119,180}    
\definecolor{phRoute}{RGB}{44,160,44}     
\definecolor{phDiss}{RGB}{214,39,40}      
\definecolor{phPort}{RGB}{255,127,14}     
\definecolor{phBand}{RGB}{148,103,189}    
\definecolor{phBG}{RGB}{247,247,247}
\definecolor{phInk}{RGB}{40,40,40}
\usepackage{cleveref}

\definecolor{hnnblue}{RGB}{0,82,155}
\definecolor{hnnorange}{RGB}{200,80,0}
\hypersetup{colorlinks=true, linkcolor=hnnblue, citecolor=hnnorange, urlcolor=hnnblue}

\theoremstyle{definition}

\theoremstyle{plain}

\newcommand{\FitTrainN}{1{,}109{,}250}
\newcommand{\FitHoldout}{S010, S011, S012}
\newcommand{\FitTestMSE}{\ensuremath{1.30\times10^{-4}}}
\newcommand{\LadderOneModel}{1.96}
\newcommand{\LadderSigmaModel}{1.00}
\newcommand{\LadderDFAModel}{1.68}
\newcommand{\LadderOneReal}{1.18}
\newcommand{\LadderSigmaReal}{0.94}
\newcommand{\LadderDFAReal}{0.68}
\newcommand{\FitTemp}{0.2}

\providecommand{\FitTrainN}{1{,}109{,}250}
\providecommand{\FitHoldout}{S010, S011, S012}
\providecommand{\FitTestMSE}{\ensuremath{1.30\times10^{-4}}}
\providecommand{\LadderOneModel}{1.96}
\providecommand{\LadderSigmaModel}{1.00}
\providecommand{\LadderDFAModel}{1.68}
\providecommand{\LadderOneReal}{1.18}
\providecommand{\LadderSigmaReal}{0.94}
\providecommand{\LadderDFAReal}{0.68}
\providecommand{\FitTemp}{0.2}

\begin{document}

\title{A Physics-Inspired Classical Digital Twin of Cortical Dynamics:\\
  A Band-Stratified Metriplectic Port-Hamiltonian Neural Network\\
  Learned from Brain--Computer-Interface EEG}

\author{Dibakar Sigdel}
\email{devdeep137@gmail.com}
\affiliation{Mindverse Computing LLC, Lynnwood, WA 98087}

\date{\today}

\begin{abstract}
We present a physics-inspired classical digital twin of brain--computer-
interface (BCI) data: a graph neural network constrained to a band-stratified,
metriplectic port-Hamiltonian form, with parameters learned from scalp EEG
recorded during rest and motor imagery. The port-Hamiltonian structure is a
modelling choice --- it buys passivity, a certified steady-state power balance,
and a clean separation of storage, routing and dissipation --- not a claim about
what the brain is. The state pairs each channel's instantaneous phase with its
angular frequency, and stored energy decomposes over the five canonical
frequency bands. A phase-locking prior measured from the same recordings gates
the learned connectome, and a metriplectic formulation places the twin at a non-
equilibrium steady state sustained by a metabolic port. Fitted to \FitTrainN\
phasor samples from the PhysioNet EEG Motor Movement/Imagery database under a
leakage-free split, the twin reaches a held-out reconstruction error of
\FitTestMSE. Scored free-running against invariants it did not author, the
verdict is mixed: it reproduces near-critical avalanche branching
($\sigma\approx1$) but not the aperiodic $1/f$ slope or the long-range temporal
correlations of the recordings. Skew-symmetry and non-negative dissipation hold
by construction rather than by penalty, making the twin a structure-preserving
substrate on which closed-loop neuromodulation can be designed and tested.
\end{abstract}


\maketitle

\section{Introduction}

Brain--computer interfacing aspires to models of recorded neural activity that are
not merely discriminative but \emph{interrogable}: models one can run forward,
perturb, and hold to account. The measurement modality that dominates non-invasive
BCI --- high-density electroencephalography (EEG) --- delivers richly structured
oscillations spanning five canonical frequency bands: delta ($\delta$, 1--4\,Hz),
theta ($\theta$, 4--8\,Hz), alpha ($\alpha$, 8--12\,Hz), beta ($\beta$, 12--30\,Hz),
and gamma ($\gamma$, 30--80\,Hz). These are not passive spectral features; they are
the scalp-level signatures of self-sustained cortical rhythms through which neural
assemblies coordinate information transfer, consolidate working memory, and regulate
attentional gating \cite{Buzsaki2006}. Turning that oscillatory record into a
governing dynamics --- one that carries an explicit energy, an explicit dissipation,
and an explicit stimulation port --- is the problem this work addresses.

\begin{figure*}[p]
  \centering
  \makebox[\textwidth][c]{\includegraphics[width=0.845\textwidth]{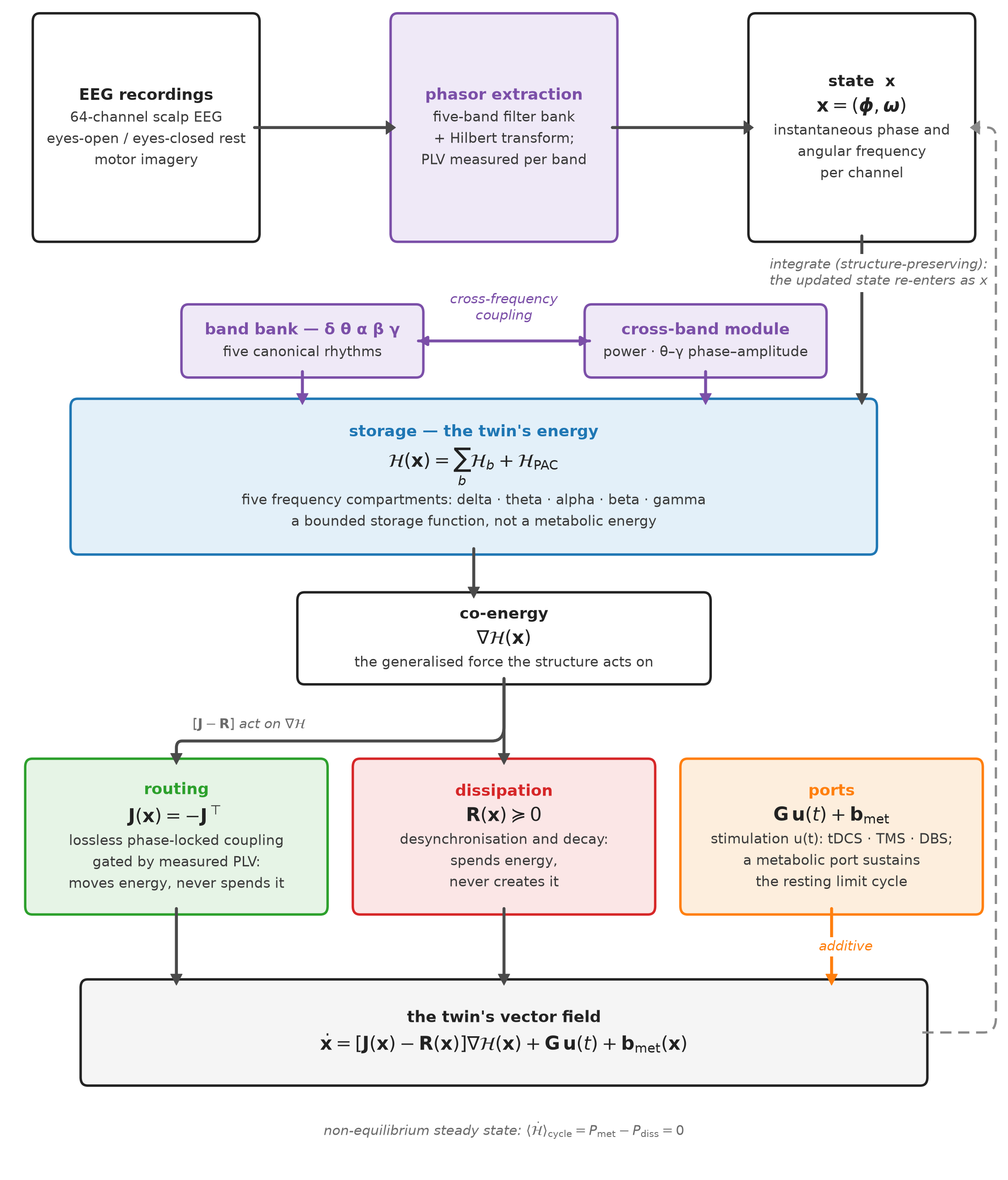}}
  \caption{\textbf{The classical port-Hamiltonian digital twin of BCI data.}
    A five-band filter bank and the Hilbert transform turn the recordings into the
    state $\mathbf{x}=(\bm{\phi},\bm{\omega})$, pairing each channel's instantaneous
    phase with its angular frequency. Storage $\mathcal{H}$ (blue) decomposes over
    the five canonical frequency compartments (purple) plus a cross-band
    phase--amplitude term; its gradient is the force on which
    $\mathbf{J}=-\mathbf{J}^\top$ (green, gated by the measured phase-locking value)
    and $\mathbf{R}\succeq 0$ (red) act, while $\mathbf{G}\mathbf{u}(t)$ (orange)
    adds stimulation and the metabolic port $\mathbf{b}_{\mathrm{met}}$ sustains the
    resting limit cycle. Skew-symmetry, positive semi-definiteness, and the
    phase-locking gate are imposed, not penalised. Every object in the diagram is a
    property of the twin, not an assertion about cortex. See
    Sec.~\ref{sec:theory}.}
  \label{fig:graphical_abstract}
\end{figure*}

The dominant paradigm for BCI signal processing relies on discriminative machine
learning architectures --- convolutional neural networks, support vector machines,
or EEGNet variants --- trained to classify categorical cognitive states from raw EEG
time-series \cite{Lotte2018}. These approaches treat the recording as an
unstructured feature stream, absorbing no structural knowledge of the dynamics that
produced it. They carry no time axis one can extrapolate, no energy budget, and no
thermodynamic constraint; consequently they cannot be asked what would happen under
a stimulation the experiment never delivered, and a representation that manufactures
energy can still score well on a classification benchmark while being impossible as
a dynamics.

A model useful for those questions needs three ingredients simultaneously: (i) a
\emph{geometry} that reflects the oscillatory, phase-locked character of the
recorded rhythms; (ii) a \emph{dynamics} that enforces thermodynamic structure,
cleanly separating conservative routing from irreversible decay; and (iii) a
\emph{surrogate} expressive enough to capture the high-dimensional, state-dependent
rewiring of functional coupling across conditions. The phasor/torus geometry
supplies (i), the port-Hamiltonian framework supplies (ii) by construction, and a
graph neural network supplies (iii).

Learning continuous-time dynamics with an embedded energy structure is by now
established: Hamiltonian Neural Networks recover conservative dynamics from data
\cite{Greydanus2019}, and dissipative extensions incorporate friction and external
control \cite{Zhong2020}. The Port-Hamiltonian (pH) framework
\cite{vanderSchaft2014} supplies the mathematical language for these requirements,
unifying conservative routing, dissipation, and actuation in a single structure
(Fig.~\ref{fig:graphical_abstract} sets out the whole construction at a glance).
The continuous-time dynamics
\begin{equation}
  \dot{\mathbf{x}} = \bigl[\mathbf{J}(\mathbf{x}) - \mathbf{R}(\mathbf{x})\bigr]\nabla_{\mathbf{x}}\mathcal{H}(\mathbf{x}) + \mathbf{G}\mathbf{u}(t)
  \label{eq:ph_dynamics}
\end{equation}
encode conservative routing ($\mathbf{J}$, skew-symmetric), irreversible dissipation ($\mathbf{R}$, positive semi-definite), and structured external intervention ($\mathbf{G}\mathbf{u}$).

The resulting power balance,
\begin{equation}
  \dot{\mathcal{H}} = -\nabla\mathcal{H}^\top \mathbf{R}\, \nabla\mathcal{H} + \mathbf{y}^\top\mathbf{u}(t) \;\leq\; \mathbf{y}^\top\mathbf{u}(t),
  \label{eq:power_balance}
\end{equation}
states that without external stimulation the twin's storage function $\mathcal{H}$ is non-increasing. Strict passivity is, however, only a first approximation for a model of waking-state recordings: it is the condition for relaxation to a fixed point, whereas the recordings show sustained oscillation, which in the brain is held far from equilibrium by continuous metabolic supply. We therefore adopt, in Section~\ref{sec:metriplectic}, a non-equilibrium steady-state (metriplectic) refinement in which an explicit metabolic port balances dissipation on the twin's resting limit cycle, and the invariant to be certified becomes orbital stability with a bounded storage function rather than monotone energy decay.

\subsection{Limitation of Static Connectomes}

A second challenge is scalability, and a failure mode of prior pHNN formulations is
the use of \emph{static} interaction matrices. Functional coupling in the recordings
is manifestly state-dependent: the same montage carries different coherence
structure at rest than under motor imagery, so a static $\mathbf{J}$ captures only
the time-averaged connectome. The $2N\times2N$ connectome $\mathbf{J}(\mathbf{x})$
must therefore be predicted as a state-dependent matrix, respecting skew-symmetry at
all states, while the dissipation matrix $\mathbf{R}(\mathbf{x})$ tracks regionally
heterogeneous decay. A static parametric matrix cannot achieve this; a graph neural
network (GNN) surrogate can. The formulation is generic in the channel count $N$;
the present experiments use $N = 64$, but nothing in the architecture is tied to
that value.

\subsection{Contribution}

We present the \textbf{Cortical GNN-pHNN}, a band-stratified, metriplectic
port-Hamiltonian twin learned by a GNN surrogate from the PhysioNet EEG Motor
Movement/Imagery database ($64$ channels, $12$ subjects, $60$ recordings). Our
contributions are:
\begin{enumerate}
  \item A \textbf{phasor state extracted from the recordings}, $\mathbf{x} =
    [\bm{\phi},\bm{\omega}]$, in which each channel's instantaneous phase is paired
    with its angular frequency, obtained by a five-band filter bank and the Hilbert
    transform rather than posited by a generator.
  \item A \textbf{band-stratified composite Hamiltonian}
    $\mathcal{H} = \sum_b \mathcal{H}_b + \mathcal{H}_{\mathrm{PAC}}$ whose
    compartments are the five canonical rhythms, augmented by a cross-band module
    that separates direct power coupling from the $\theta$--$\gamma$
    phase--amplitude coupling associated with working-memory indexing, so the
    twin's energy decomposition is readable band by band.
  \item A \textbf{phase-locking-gated dynamic connectome}: the routing operator is
    anti-symmetrised exactly at every state and multiplicatively gated by the
    phase-locking values measured on the same recordings, so coupling is admitted
    only where coherence is present rather than invented freely.
  \item A \textbf{state-dependent dissipation} $\mathbf{R}(\mathbf{x}) =
    \mathrm{diag}(r_j(\mathbf{x})) \succeq 0$ enforced by construction, capturing
    regionally heterogeneous decay without a positivity penalty.
  \item A \textbf{metriplectic non-equilibrium reformulation}
    (Section~\ref{sec:metriplectic}) in which reversible and irreversible generators
    coexist under exact degeneracy, an explicit metabolic port sustains the twin's
    resting limit cycle, and a fluctuation--dissipation-consistent noise channel is
    governed by a single arousal temperature.
  \item An \textbf{honest scoring against invariants the twin did not author}:
    a falsifiable ladder --- $1/f$ spectral fidelity, functional connectivity,
    criticality (avalanche branching and long-range temporal correlations) ---
    measured identically on the recordings and on the twin's free run, reporting
    which rungs it clears (branching) and which it does not (spectral slope, DFA),
    with each gap traced to a specific, testable upgrade.
\end{enumerate}

Structured deformations of this reference twin --- the frame in which subject
personalisation and neurological disorder are expressed --- are developed and
proved in the companion theory manuscript, and applied in the companion atlas and
diseases manuscripts; the present work fixes the reference twin they deform.

\paragraph{Positioning.} The framework occupies ground that neither established
camp reaches alone. Unlike discriminative BCI decoders \cite{Lotte2018}, which map
a window of EEG to a label, it is a continuous-time \emph{dynamics} with an explicit
energy, dissipation, and stimulation port --- so passivity and the steady-state
power balance hold by construction rather than being hoped for, and the twin can be
asked about conditions the experiment did not visit. Unlike hand-built neural-mass
and neural-field models \cite{JansenRit1995,WongWang2006,SanzLeon2013}, whose
parameters are curated per mechanism, its connectome and dissipation are learned
end-to-end from the recordings. And unlike generic structure-preserving networks
\cite{Greydanus2019,Zhong2020}, whose conserved quantity is an abstract learned
scalar, its storage function is decomposed over named frequency compartments and
its ports are mapped to physical stimulation modalities, so the model class
--- and its failures --- can be interrogated object by object.

\paragraph{What this model is, and what it is not.}
The object developed here is a \emph{physics-inspired classical digital twin} of
brain--computer-interface data: a neural network whose architecture is constrained
to the port-Hamiltonian form and whose parameters are learned from EEG
measurements. The port-Hamiltonian structure is a modelling choice, adopted because
it supplies passivity, a certified power balance and a clean separation of storage,
routing and dissipation --- properties that make a learned model interrogable and
its failures diagnosable. It is not a claim about what the brain is.

The brain is
open, stochastic, spatially organised, and almost certainly not exactly Hamiltonian
at any level of description; the value of the present construction is that it
\emph{mimics} measured cortical activity well enough to be probed under conditions
the experiment did not visit, while carrying physical guarantees by construction
rather than by penalty. The twin is moreover fitted to \emph{scalp} EEG, a low-rank
mixed projection of the underlying field, so what it tracks is the recorded signal
and the dynamical invariants that signal carries --- not a subject's cortical
anatomy. Every result below should be read as a statement about the twin, and about
how faithfully the twin tracks the data --- never as a statement that the brain
obeys these equations.

\paragraph{A note on the word ``cortex''.}
Throughout this paper, unless the context is explicitly neurophysiological,
\emph{cortex} means the classical port-Hamiltonian digital twin of the recorded
cortical activity --- the learned model, not the living brain. This convention lets
the exposition read naturally without repeating the qualifier: statements about
``the cortex's'' energy, connectome or operating point are statements about the
twin. 

Where a sentence concerns the brain itself, or the recordings, it says so.

\section{Theory: A Band-Stratified Metriplectic Twin}\label{sec:theory}

\subsection{Neural Oscillators as Stuart-Landau Limit Cycles}

Neural assemblies sustain rhythmic activity through the competition between amplifying excitatory recurrence and inhibitory feedback. In the vicinity of a Hopf bifurcation --- the standard transition point for cortical rhythm generation --- the band-limited activity of channel $j$ in frequency band $b$ admits a universal reduced description, the \textbf{Stuart-Landau} (complex Ginzburg-Landau) normal form \cite{Strogatz2018}, which is the motivation for the twin's phasor coordinates:
\begin{equation}
  \dot{z}_{j}^{(b)} = \left(\lambda_b + i\omega_{j}^{(b)}\right)z_{j}^{(b)} - \beta_{j}^{(b)}\left|z_{j}^{(b)}\right|^2 z_{j}^{(b)} + \kappa_{j}^{(b)},
  \label{eq:sl_oscillator}
\end{equation}
where $\lambda_b > 0$ is the band-specific growth rate, $\omega_{j}^{(b)} \in [\omega_b^{\min}, \omega_b^{\max}]$ is the natural frequency drawn from the band's spectral range, $\beta_{j}^{(b)} > 0$ is the nonlinear saturation coefficient, and $\kappa_{j}^{(b)}$ encodes cross-channel coupling drives. At steady state, $|z_{j}^{(b)}| = \sqrt{\lambda_b / \beta_{j}^{(b)}}$, consistent with empirically observed EEG amplitude statistics.

The \textbf{canonical phasor state} of channel $j$ is:
\begin{equation}
  z_j = A_j(t)\,e^{i\phi_j(t)},
\end{equation}
extracted from the broadband EEG signal via Butterworth bandpass filtering followed by the Hilbert transform. The instantaneous amplitude $A_j$ and unwrapped phase $\phi_j$ are the biologically meaningful coordinates. Defining the angular frequency $\omega_j = \dot{\phi}_j$, the \textbf{canonical Hamiltonian state vector} is:
\begin{equation}
  \mathbf{x}(t) = [\phi_1, \ldots, \phi_N, \omega_1, \ldots, \omega_N]^\top \in \mathbb{R}^{2N},\quad N = 64,
\end{equation}
which maps the entire cortical state onto the $N$-dimensional torus $\mathbb{T}^N$ \cite{Kuramoto1984}.

\subsection{Cognitive States, Band Dynamics, and Scalp Topology}

The five canonical EEG bands index qualitatively distinct cortical mechanisms. Delta ($\delta$, 1--4\,Hz) accompanies deep rest, slow-wave sleep consolidation, and fatigue recovery; theta ($\theta$, 4--8\,Hz) supports hippocampal memory indexing and prefrontal working-memory maintenance; alpha ($\alpha$, 8--12\,Hz) reflects idling and inhibitory gating of posterior cortex and is suppressed under cognitive load; beta ($\beta$, 12--30\,Hz) subserves sensorimotor integration and the active maintenance of motor set; and gamma ($\gamma$, 30--80\,Hz) carries fast cortical computation, perceptual binding, and task-relevant oscillations.

The recorded EEGMMIDB conditions exercise these mechanisms directly: eyes-open
and eyes-closed rest differ in posterior alpha (the classic Berger effect --- eye
closure enhances occipital $\alpha$), while the motor-imagery conditions engage
sensorimotor $\beta$ and event-related desynchronisation. The cognitive-state
phenomenology above is thus present in the data itself rather than imposed by a
generator.

The 64 channels are anatomically organised into four scalp regions --- Frontal (F, channels 1--16), Temporal (T, 17--32), Parietal (P, 33--48), and Occipital (O, 49--64) --- each with neuroanatomically grounded cross-regional coupling:
\begin{equation}
  \kappa_{j}^{(\theta)} = \epsilon_\theta \cdot \bar{z}_\mathrm{Frontal}^{(\theta)}, \quad j \in \mathrm{Parietal},
  \label{eq:fp_coupling}
\end{equation}
encoding the well-established \textbf{fronto-parietal theta coherence network} active during working memory \cite{Sauseng2005}.

\subsection{Phase Locking Value and the Coupling Prior}

The \textbf{Phase Locking Value} (PLV) between channels $j$ and $k$ in band $b$ is the empirical magnitude of the mean complex phase difference:
\begin{equation}
  \mathrm{PLV}_{jk}^{(b)} = \left|\frac{1}{T}\sum_{t=1}^T e^{i\bigl(\phi_j^{(b)}(t) - \phi_k^{(b)}(t)\bigr)}\right| \in [0,1].
  \label{eq:plv}
\end{equation}
$\mathrm{PLV}_{jk} = 1$ indicates perfect phase locking (coherent coupling); $\mathrm{PLV}_{jk} = 0$ indicates phase incoherence. Cross-band PLV (e.g., $\mathrm{PLV}_{\theta\gamma}$) approximates Phase-Amplitude Coupling (PAC) strength when the gamma envelope is modulated by theta phase --- the canonical hippocampal working-memory indexing mechanism \cite{Lisman2013}.

These empirical PLV matrices serve as \emph{biologically principled priors} for the learned coupling matrix $\mathbf{J}(\mathbf{x})$, enforcing that the twin's connectome reflects measured synchrony structure.

\subsection{Port-Hamiltonian Structure and Thermodynamic Stability}

The twin's dynamics (Eq.~\ref{eq:ph_dynamics}) decompose into three structurally distinct terms \cite{vanderSchaft2014}. The skew-symmetric \textbf{functional connectome} $\mathbf{J}(\mathbf{x}) = -\mathbf{J}(\mathbf{x})^\top$ carries conservative, lossless phase-locked routing: a non-zero entry $J_{jk}(\mathbf{x})$ means that the twin's channels $j$ and $k$ exchange stored energy without dissipating it, the model's counterpart of the fronto-parietal alpha coherence seen in attentional engagement. The positive semi-definite \textbf{dissipation matrix} $\mathbf{R}(\mathbf{x}) \succeq 0$ carries irreversible loss, so that a large regional value $r_j(\mathbf{x})$ marks local desynchronisation --- the model's counterpart of the loss of sustained rhythmic activity reported in cognitive fatigue, ADHD, or seizure termination. The \textbf{neuromodulation port} $\mathbf{G}\mathbf{u}(t)$ injects energy through three structurally distinct channels standing in for transcranial direct-current stimulation (tDCS) of prefrontal circuitry, transcranial magnetic stimulation (TMS) of the dorsal attention network, and deep-brain stimulation (DBS) or theta entrainment of hippocampal--entorhinal rhythms. These identifications are how stimulation modalities \emph{enter the model}; whether the ports are physical rather than decorative is a question the twin must answer under an actively delivered perturbation, which is rung~5 of the validation ladder below.

The \textbf{passivity invariant} (Eq.~\ref{eq:power_balance}) has a direct reading: with no stimulation ($\mathbf{u} = 0$), the twin cannot spontaneously amplify its storage function without bound. A gross violation would mean a model predicting infinite-energy oscillations, which is a defect of the model rather than a discovery about the brain. We stress, however, that strict global passivity ($\dot{\mathcal{H}} \leq 0$ everywhere) is only a \emph{first approximation} for a model of waking-state recordings: it is the condition for relaxation to a fixed point, whereas the recordings show sustained oscillation --- in the brain, maintained far from equilibrium by continuous metabolic supply. Section~\ref{sec:metriplectic} refines the passivity picture into a non-equilibrium steady-state (NESS) formulation that resolves this tension and defines the correct invariant to track.

\subsection{From Passivity to a Non-Equilibrium Steady State: A Metriplectic Formulation}
\label{sec:metriplectic}

The Stuart-Landau normal form (Eq.~\ref{eq:sl_oscillator}) contains an \emph{active} term: for $\lambda_b > 0$ the origin is unstable, so the oscillator injects energy at low amplitude and only saturates at the limit cycle $|z| = \sqrt{\lambda_b/\beta}$. A strictly passive autonomous port-Hamiltonian system, by contrast, monotonically dissipates its storage function and relaxes to the energy minimum --- a twin that falls silent. Reconciling recordings whose reduced description is an \emph{active} limit cycle with a model certified as globally \emph{passive} therefore requires a structure that supports a persistent, energy-consuming steady state rather than a decaying one. The physical picture the twin borrows is that the waking brain does not sit at an energy minimum but is maintained at a \textbf{non-equilibrium steady state} by a continuous metabolic energy flux that balances dissipation \cite{Deco2011,Breakspear2017}; the twin imports this as a structural requirement on its own vector field, not as a measurement.

The natural generalisation is the \textbf{metriplectic} (or GENERIC) bracket formalism \cite{Morrison1986,GrmelaOttinger1997,Ottinger2005}, which augments the conservative Poisson structure with an explicit irreversible, entropy-generating bracket. The dynamics carry two potentials --- the regulatory energy $\mathcal{H}(\mathbf{x})$ and a cortical \textbf{entropy} (disorder) functional $\mathcal{S}(\mathbf{x})$ ---
\begin{equation}
  \dot{\mathbf{x}} = \underbrace{\mathbf{J}(\mathbf{x})\,\nabla \mathcal{H}(\mathbf{x})}_{\text{reversible routing}} \;+\; \underbrace{\mathbf{M}(\mathbf{x})\,\nabla \mathcal{S}(\mathbf{x})}_{\text{irreversible production}} \;+\; \underbrace{\mathbf{G}\,\mathbf{u}(t)}_{\text{stimulation port}} \;+\; \underbrace{\mathbf{b}_{\mathrm{met}}(\mathbf{x})}_{\text{metabolic port}},
  \label{eq:metriplectic}
\end{equation}
with $\mathbf{M}(\mathbf{x}) = \mathbf{M}(\mathbf{x})^\top \succeq 0$ and the \emph{degeneracy conditions} $\mathbf{J}\nabla\mathcal{S} = \mathbf{0}$ and $\mathbf{M}\nabla\mathcal{H} = \mathbf{0}$. These conditions decouple the two brackets and yield, for the autonomous system ($\mathbf{u}=0$, $\mathbf{b}_{\mathrm{met}}=0$),
\begin{equation}
  \dot{\mathcal{H}} = \nabla\mathcal{H}^\top \mathbf{J}\, \nabla\mathcal{H} = 0
  \qquad\text{and}\qquad
  \dot{\mathcal{S}} = \nabla\mathcal{S}^\top \mathbf{M}\, \nabla\mathcal{S} \;\geq\; 0,
  \label{eq:entropy_production}
\end{equation}
i.e.\ exact energy conservation of the reversible part together with a manifest \textbf{second law} (non-negative entropy production). The dissipation matrix of the pH form is recovered as $\mathbf{R} = -\mathbf{M}\,\nabla^2\mathcal{S}\,(\nabla\mathcal{H})^{+}$ near the operating point, so the state-dependent dissipation of the present formulation is the linearisation of the metriplectic irreversible bracket.

In this picture the twin's resting state is characterised not by $\dot{\mathcal{H}} \leq 0$ but by a \textbf{steady-state power balance} on the limit cycle,
\begin{equation}
  \big\langle \dot{\mathcal{H}} \big\rangle_{\mathrm{cycle}} = P_{\mathrm{met}} - P_{\mathrm{diss}} = 0,
  \qquad P_{\mathrm{diss}} = \big\langle \nabla\mathcal{H}^\top \mathbf{R}\,\nabla\mathcal{H}\big\rangle \geq 0,
  \label{eq:ness_balance}
\end{equation}
where the metabolic port $P_{\mathrm{met}}$ replaces the energy dissipated per cycle. The correct stability notion is therefore \emph{orbital} (Floquet) stability of the limit cycle with a storage function that is bounded below, not monotone decrease of $\mathcal{H}$ toward zero. Within the twin, pathology is represented as a \emph{failure of this balance}: runaway entropy production and collapse of $P_{\mathrm{met}}$, the model's analogue of fatigue, anaesthesia or coma, or, conversely, a dissipation failure driving hypersynchronous energy growth as an analogue of seizure onset. These are correspondences the model offers for testing, not mechanisms it demonstrates.

Finally, the metriplectic form makes fluctuations thermodynamically consistent. Adding a stochastic drive whose covariance is tied to the irreversible bracket through the fluctuation--dissipation theorem \cite{Kubo1966},
\begin{equation}
  d\mathbf{x} = \big[\mathbf{J}\nabla\mathcal{H} + \mathbf{M}\nabla\mathcal{S} + \mathbf{G}\mathbf{u}\big]\,dt + \boldsymbol{\sigma}(\mathbf{x})\,d\mathbf{W},
  \qquad \boldsymbol{\sigma}\boldsymbol{\sigma}^\top = 2\,T\,\mathbf{R}(\mathbf{x}),
  \label{eq:fdt}
\end{equation}
introduces a single scalar \textbf{temperature} $T$ --- a proxy for neuromodulatory arousal --- along which discrete cognitive conditions become locations on a continuous arousal axis. This stochastic, metriplectic extension is the theoretical backbone of the upgrade programme developed in Section~\ref{sec:upgrade}.

\section{Methods: The Cortical GNN-pHNN Architecture}
\label{sec:methods}

\subsection{EEG data and preprocessing}

We use cortical recordings from the PhysioNet EEG Motor Movement/Imagery
database (EEGMMIDB;~\cite{schalk2004,goldberger2000}): $N=64$ channels recorded at
$160$\,Hz on the international 10--10 montage, $12$ subjects, $60$ recordings
spanning eyes-open rest, eyes-closed rest, and two motor-imagery paradigms
(left/right hand; hands/feet). Rest recordings run $\sim$61\,s and task recordings
$\sim$123--125\,s. Each recording is average-referenced,
band-pass filtered ($1$--$45$\,Hz), resampled to $f_s=250$\,Hz, and per-channel
$z$-scored; the five canonical bands are then separated by fourth-order
Butterworth filters ($\delta$ 1--4, $\theta$ 4--8, $\alpha$ 8--12,
$\beta$ 12--30, $\gamma$ 30--80\,Hz). From each band a Hilbert transform yields the instantaneous
phase $\phi_j^{(b)}(t)$ and angular frequency $\omega_j^{(b)}(t)$ that make up the
canonical Hamiltonian coordinates $\mathbf{x}=[\bm{\phi},\bm{\omega}]$; the
empirical phase-locking-value matrices are computed on the same bands and supply
the coherence prior of Eq.~\ref{eq:plv_gate}. The Stuart-Landau oscillator of
Section~\ref{sec:metriplectic} (Eq.~\ref{eq:sl_oscillator}) is the theoretical
normal form that motivates these phasor coordinates near a Hopf bifurcation; it is
not used to generate the data. The architecture below is written for arbitrary
$N$.

\paragraph{Phasor extraction.}
Each band $b \in \{\delta, \theta, \alpha, \beta, \gamma\}$ is isolated with a fourth-order Butterworth bandpass filter over $[\omega_b^{\min}, \omega_b^{\max}]$, and the corresponding analytic signal $z_j^{(b)}(t) = x_j^{(b)}(t) + i\,\mathcal{H}\{x_j^{(b)}\}(t)$ is obtained by the Hilbert transform. From it we extract the unwrapped instantaneous phase $\phi_j^{(b)} = \arg(z_j^{(b)})$ and the angular frequency $\omega_j^{(b)} = \dot{\phi}_j^{(b)}$, mask channels whose median amplitude falls below $5\%$ of the band maximum, and assemble the phase-locking-value matrix $\mathbf{PLV}^{(b)} \in [0,1]^{N\times N}$ (Eq.~\ref{eq:plv}). The canonical state vector $\mathbf{x}(t) \in \mathbb{R}^{128}$ is formed from the alpha-band phase and frequency as the primary phasor coordinate.

\subsection{Band-Stratified Graph Energy Network}

\begin{figure*}[p]
  \centering
  \includegraphics[width=\textwidth]{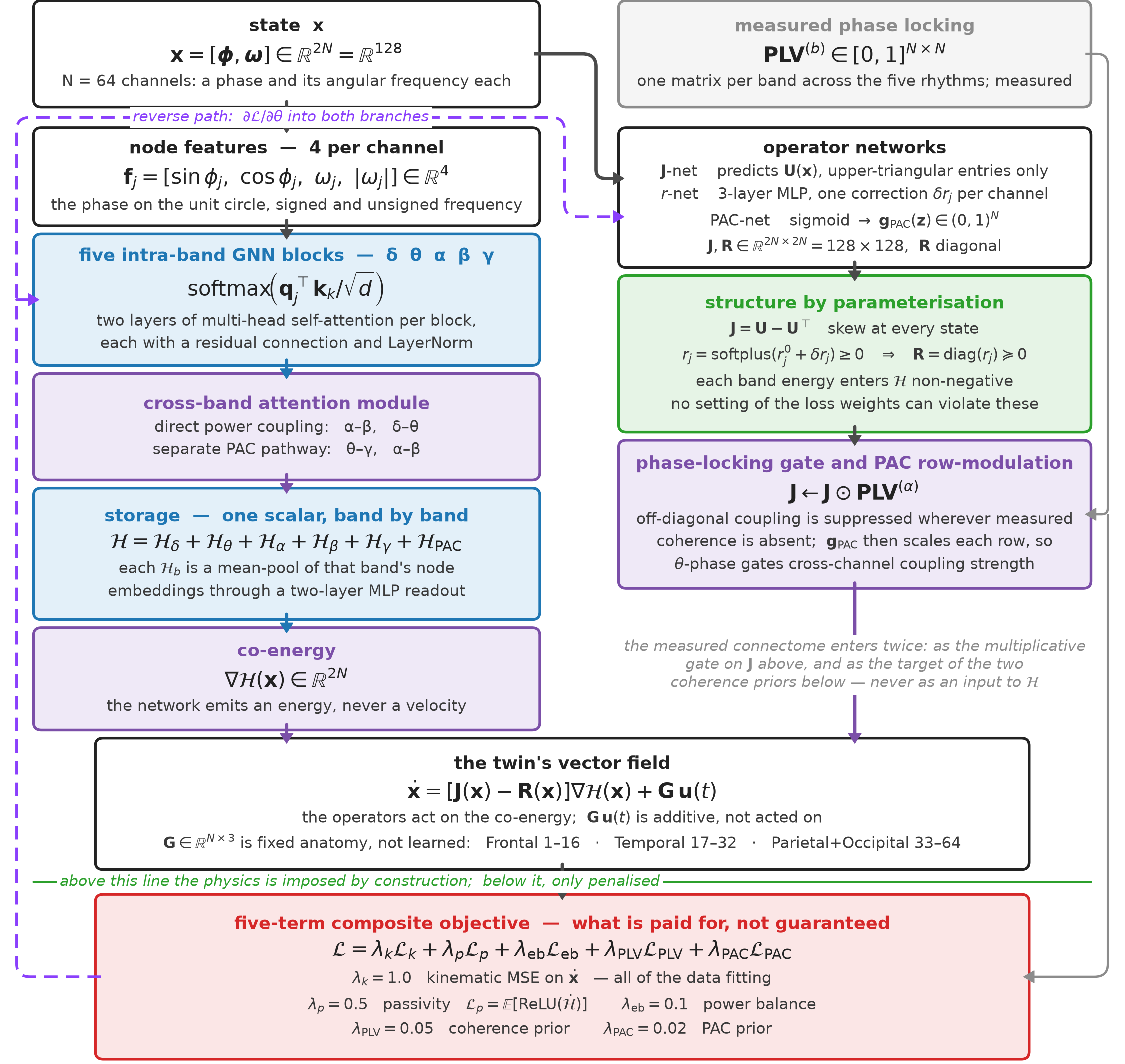}
  \caption{\textbf{Architecture of the cortical GNN-pHNN: the division of labour
  between the networks, and which properties are structural.} The left column is
  the only scalar-valued branch --- five intra-band graph blocks and a cross-band
  module funnel into one number, the storage $\mathcal{H}$, whose \emph{gradient}
  is the force the dynamics act on; the right column carries the operators and
  never emits an energy. The horizontal rule is the figure's point. Above it the
  properties are parameterisations --- skew-symmetry from
  $\mathbf{J}=\mathbf{U}-\mathbf{U}^\top$, non-negative dissipation from the
  softplus --- which no setting of the weights can violate. Below it they are
  prices that training can trade against the kinematic fit, so a residual
  violation is a quantity to report rather than a contradiction. Every quantity
  shown is a property of the twin, not an assertion about cortex.}
  \label{fig:architecture}
\end{figure*}

Figure~\ref{fig:architecture} lays out the whole construction and the boundary
that organises it: the properties drawn above the rule hold by parameterisation,
while those below are terms in the objective. The Hamiltonian $\mathcal{H}(\mathbf{x})$ is parameterised by a hierarchical GNN with five intra-band blocks and a cross-band attention module. Node features for channel $j$ are:
\begin{equation}
  \mathbf{f}_j = \bigl[\sin(\phi_j),\; \cos(\phi_j),\; \omega_j,\; |\omega_j|\bigr] \in \mathbb{R}^4,
\end{equation}
encoding the phase on the unit circle together with the signed and unsigned instantaneous frequency.

Each \textbf{intra-band GNN block} consists of two layers of multi-head self-attention with residual connections and LayerNorm, computing:
\begin{equation}
  \mathbf{h}^{(b)} = \mathrm{LayerNorm}\!\left(\mathbf{h}_0^{(b)} + \sum_{k}\alpha_{jk}^{(b)}\mathbf{W}_v^{(b)}\mathbf{h}_k^{(b)}\right),
  \quad \alpha_{jk}^{(b)} = \frac{\exp\!\left(\mathbf{q}_j^{(b)\top}\mathbf{k}_k^{(b)}/\sqrt{d}\right)}{\sum_{l}\exp\!\left(\mathbf{q}_j^{(b)\top}\mathbf{k}_l^{(b)}/\sqrt{d}\right)},
  \label{eq:self_attn}
\end{equation}
with $d = \dim(\mathbf{h})$. The attention weights $\alpha_{jk}^{(b)}$ form a dynamically learned adjacency matrix over the 64 channels within each band.

The \textbf{cross-band attention module} implements two biologically distinct coupling mechanisms. Direct power coupling --- the mass-bond analogue in Port-Hamiltonian terminology --- lets the $\alpha$--$\beta$ and $\delta$--$\theta$ pairs exchange energy through standard cross-attention between band embeddings. Phase-amplitude coupling, by contrast, is carried on a separate modulated pathway in which the $\theta$--$\gamma$ and $\alpha$--$\beta$ pairs encode the hippocampal $\theta$-phase gating of $\gamma$-amplitude that underlies working-memory indexing \cite{Lisman2013}.

The total energy decomposes as:
\begin{equation}
  \mathcal{H}(\mathbf{x}) = \mathcal{H}_\delta + \mathcal{H}_\theta + \mathcal{H}_\alpha + \mathcal{H}_\beta + \mathcal{H}_\gamma + \mathcal{H}_{\mathrm{PAC}},
  \label{eq:energy_decomp}
\end{equation}
where each $\mathcal{H}_b$ is computed by mean-pooling the intra-band node embeddings and passing through a two-layer MLP readout. The decomposition is what makes the twin's energy readable: the sub-energies can be inspected against the band phenomenology of the recorded conditions --- posterior $\mathcal{H}_\alpha$ dominance in eyes-closed rest, and sensorimotor $\mathcal{H}_\beta$ modulation under motor imagery --- so a mismatch is visible band by band rather than hidden in one scalar.

\subsection{Phase-Locking-Gated Dynamic Connectome}

The dynamic functional connectome $\mathbf{J}(\mathbf{x})$ is predicted by a dedicated network operating on the encoded state. A shared encoder $\mathbf{z} = f_\phi(\mathbf{x}) \in \mathbb{R}^d$ first extracts a latent cortical representation, from which the upper-triangular elements $\{J_{jk}\}_{j<k}$ are predicted and the full matrix is anti-symmetrised exactly,
\begin{equation}
  \mathbf{J}(\mathbf{x}) = \mathbf{U}(\mathbf{x}) - \mathbf{U}(\mathbf{x})^\top,\quad \mathbf{U}(\mathbf{x})_{jk} = 0 \text{ for } j \geq k,
  \label{eq:skew_sym}
\end{equation}
so that skew-symmetry holds at every state by construction. The alpha-band phase-locking matrix then acts as a multiplicative coupling prior,
\begin{equation}
  \mathbf{J}(\mathbf{x}) \leftarrow \mathbf{J}(\mathbf{x}) \odot \mathbf{PLV}^{(\alpha)},
  \label{eq:plv_gate}
\end{equation}
suppressing off-diagonal coupling wherever empirical phase coherence is absent. Finally, a per-channel amplitude-modulation vector $\mathbf{g}_\mathrm{PAC}(\mathbf{z}) \in (0,1)^N$, produced by a sigmoid network and applied row-wise, enforces the theta-phase gating of cross-channel coupling strength.

\subsection{State-Dependent Dissipation}

Per-channel dissipation is modelled as:
\begin{equation}
  r_j(\mathbf{x}) = \mathrm{softplus}\!\left(r_j^0 + \delta r_j(\mathbf{x})\right) \geq 0,
  \label{eq:r_net}
\end{equation}
where $r_j^0$ is a learned baseline decay rate, standing in for the myelination and synaptic time constant, and $\delta r_j(\mathbf{x})$ is a state-dependent correction estimated by a three-layer MLP. The softplus activation guarantees $\mathbf{R}(\mathbf{x}) = \mathrm{diag}(r_j(\mathbf{x})) \succeq 0$ at all states. The quantity is readable against the data: elevated regional $r_j$ in the twin marks local desynchronisation --- for example the occipital dissipation that should accompany alpha suppression on eye opening.

\subsection{Neuroanatomical Port Matrix and Neuromodulation}

The port matrix $\mathbf{G} \in \mathbb{R}^{N \times 3}$ is initialised with neuroanatomical structure:
\begin{equation}
  \begin{aligned}
  G_{j,0} &= \mathbf{1}[j \in \mathrm{Frontal}]/|\mathrm{Frontal}|,\quad
  G_{j,1} = \mathbf{1}[j \in \mathrm{P}\cup\mathrm{O}]/|\mathrm{P}\cup\mathrm{O}|,\\
  G_{j,2} &= \mathbf{1}[j \in \mathrm{Temporal}]/|\mathrm{Temporal}|,
  \end{aligned}
  \label{eq:G_matrix}
\end{equation}
where $\mathrm{Frontal}$ (indices 1--16), Parietal+Occipital (33--64), and $\mathrm{Temporal}$ (17--32) correspond to standard 10-20 electrode groupings. Port 0 stands in for tDCS to the prefrontal cortex; Port 1 for TMS to the dorsal attention network; Port 2 for hippocampal theta entrainment via DBS or transcranial alternating current stimulation (tACS). This is a coarse anatomical averaging that fixes \emph{where} each modality enters the twin; replacing it with a head-model-projected lead field is one of the upgrades of Section~\ref{sec:upgrade}.

\subsection{Five-Term Physics-Informed Composite Loss}

Training minimises:
\begin{equation}
  \mathcal{L} = \lambda_k \mathcal{L}_k + \lambda_p \mathcal{L}_p + \lambda_\mathrm{eb} \mathcal{L}_\mathrm{eb} + \lambda_\mathrm{PLV} \mathcal{L}_\mathrm{PLV} + \lambda_\mathrm{PAC} \mathcal{L}_\mathrm{PAC},
  \label{eq:composite_loss}
\end{equation}
with weights $\lambda_k\!=\!1.0$, $\lambda_p\!=\!0.5$, $\lambda_\mathrm{eb}\!=\!0.1$, $\lambda_\mathrm{PLV}\!=\!0.05$, $\lambda_\mathrm{PAC}\!=\!0.02$. These five terms are the
penalised half of Figure~\ref{fig:architecture}: unlike the skew-symmetry of
$\mathbf{J}$ and the non-negativity of the dissipation rates, none of them is
guaranteed by the parameterisation, so each is a property the fit can trade away
and must therefore be reported rather than assumed.

\paragraph{Kinematic accuracy ($\mathcal{L}_k$).}
$\mathcal{L}_k = \|\dot{\mathbf{x}}_\mathrm{pred} - \dot{\mathbf{x}}_\mathrm{true}\|^2$ ensures the model reproduces the observed cortical trajectories.

\paragraph{Passivity regulariser ($\mathcal{L}_p$).}
$\mathcal{L}_p = \mathbb{E}\!\left[\mathrm{ReLU}(\dot{\mathcal{H}})\right]$ penalises any positive energy increase at zero stimulation, enforcing the passivity invariant.

\paragraph{Energy balance ($\mathcal{L}_\mathrm{eb}$).}
$\mathcal{L}_\mathrm{eb} = \left|\dot{\mathcal{H}} + \nabla\mathcal{H}^\top \mathbf{R}\nabla\mathcal{H} - \mathbf{y}^\top\mathbf{u}\right|^2$ enforces exact consistency with the Port-Hamiltonian power balance equation.

\paragraph{PLV coherence prior ($\mathcal{L}_\mathrm{PLV}$).}
$\mathcal{L}_\mathrm{PLV} = \left\|\widetilde{|\mathbf{J}|} - \widetilde{\mathbf{PLV}^{(\alpha)}}\right\|_F^2$ (where $\tilde{\cdot}$ denotes normalisation to $[0,1]$) anchors the coupling magnitude to empirically measured synchrony.

\paragraph{PAC regulariser ($\mathcal{L}_\mathrm{PAC}$).}
$\mathcal{L}_\mathrm{PAC} = \left\|\widetilde{|\mathbf{J}_{\theta\gamma}|} - \widetilde{\mathbf{PLV}^{(\theta\gamma)}}\right\|_F^2$ specifically guides the theta--gamma cross-band coupling block to encode the hippocampal PAC structure.

Training uses the Adam optimiser with cosine annealing ($\eta_\mathrm{max} = 5\times10^{-4}$, $\eta_\mathrm{min} = 10^{-5}$) over a cosine schedule of up to 400 epochs, with gradient clipping at $\|\cdot\| = 2.0$ for stability.

\subsection{Fitting protocol and data handling}
\label{sec:fit_protocol}

The phasor coordinates and their time derivatives are extracted once for all
$60$ recordings and concatenated into a single design matrix
$\mathbf{X}\in\mathbb{R}^{T\times128}$ with $T\approx1.48\times10^{6}$ samples,
together with the matching derivative targets $\dot{\mathbf{X}}$. The split is
\textbf{leakage-free by subject}: three subjects (S010, S011, S012) are held out
entirely for testing and contribute no sample to training, giving
$\FitTrainN$ training and $368{,}250$ test samples. We fit three random seeds; on
the held-out subjects the model reaches a kinematic reconstruction MSE of
$\FitTestMSE$ (mean over seeds), and the training curves (Fig.~\ref{fig:training_loss})
converge to a common loss floor across seeds, apart from occasional transient
optimiser excursions in one seed that recover within a few epochs. The fitted single-seed model is used for the
free-running validation of Section~\ref{sec:ladder_results}.

\subsection{Model-independent validation diagnostics}
\label{sec:diagnostics}

The validation ladder scores the free-running model against three invariants that
the model did not author, each estimated by a standard, parameter-light procedure.
\emph{(i)~Aperiodic $1/f$ slope $\beta$}: the Welch power spectral density is
computed per channel, averaged, and a line is fit to $\log P$ versus $\log f$ over
$2$--$45$\,Hz; $\beta$ is the negative slope. \emph{(ii)~Avalanche branching
parameter $\sigma$}: the multichannel signal is thresholded at $2.5$ standard
deviations to define neuronal avalanches, and $\sigma$ is estimated as the mean
ratio of descendant to ancestor events across successive time bins, with
$\sigma\!=\!1$ marking the critical point. \emph{(iii)~Detrended fluctuation
analysis exponent $\alpha$}: the root-mean-square fluctuation of the
cumulative-sum profile is regressed against window size in log--log coordinates;
$\alpha\in[0.5,1]$ indicates persistent long-range temporal correlations. The same
three estimators are applied identically to the recordings (to set the
targets, Table~\ref{tab:real_targets}) and to the model's free-run output (to score
it), so the comparison is like-for-like.

\subsection{State Dimensions, Parameter Budget, and Computational Complexity}
\label{sec:complexity}

For reproducibility, and to locate precisely where a full whole-brain model would
become expensive, it is worth stating the sizes of the objects the model carries.
The model works in \emph{vector and matrix} form throughout: no higher-order tensor
state is introduced.

\paragraph{State and operator dimensions.} Each of the $N{=}64$ EEG channels carries
an instantaneous phase $\phi_j$ and its angular frequency $\omega_j$, so the
dynamical state is a real vector $\mathbf{x}=[\bm{\phi},\bm{\omega}]\in\mathbb{R}^{2N}
=\mathbb{R}^{128}$. The interconnection and dissipation
operators are real matrices $\mathbf{J}(\mathbf{x}),\mathbf{R}(\mathbf{x})\in
\mathbb{R}^{2N\times2N}$ ($128\times128$): $\mathbf{J}$ is skew and gated to the
empirical phase-locking connectome (only $\mathcal{O}(N+|E|)$ effective couplings),
while $\mathbf{R}=\mathrm{diag}(r_j)\succeq0$ is diagonal with $\mathcal{O}(N)$
nonzeros. The neuromodulation port matrix is $\mathbf{G}\in\mathbb{R}^{N\times3}$
(three anatomical ports). Per-band phase-locking priors enter as matrices
$\mathbf{PLV}^{(b)}\in[0,1]^{N\times N}$ across the five bands. The storage function
$\mathcal{H}:\mathbb{R}^{2N}\to\mathbb{R}$ is a scalar; its gradient
$\nabla\mathcal{H}\in\mathbb{R}^{2N}$ is the field the dynamics integrate.

\paragraph{Parameter budget.} The trainable parameters are dominated not by the
operators but by the hierarchical GNN energy surrogate that produces $\mathcal{H}$
--- five intra-band blocks plus the cross-band attention module --- together with
the connectome-predictor network for $\mathbf{J}(\mathbf{x})$; the $64$-channel
montage model carries on the order of $1.8\times10^6$ parameters. The
physically-structured pieces are comparatively tiny: the diagonal dissipation rates,
the three-column port matrix, and the PLV gating together scale only as
$\mathcal{O}(N+|E|)$. The parameter count is therefore \emph{not} quadratic in $N$:
it is a fixed-size surrogate core plus a term linear in the functional graph. (The
$\mathcal{O}(N^2)$ figure below is the \emph{dense-operator storage}, a memory cost,
not a trainable-parameter count --- the two should not be conflated.)

\paragraph{Computational complexity.} Assembling and storing the dense operators is
$\mathcal{O}(N^2)$ in memory ($\sim\!1.6\times10^4$ entries at $N{=}64$), and each
forward step --- evaluating $\nabla\mathcal{H}$, forming
$(\mathbf{J}-\mathbf{R})\nabla\mathcal{H}$, and integrating --- is $\mathcal{O}(N^2)$
per step in the dense implementation (reducible to $\mathcal{O}(N+|E|)$ via the
sparse PLV gating). Every column here is \emph{polynomial}, and the classical model
runs comfortably at the full $64$-channel montage on a workstation.

\paragraph{Where this becomes the bottleneck.} The polynomial cost is benign for a
single montage but grows with the size of the \emph{explicit} representation. A
high-density montage ($N\sim256$ channels) already pushes the dense
$2N\times2N$ operators to $\mathcal{O}(N^2)\sim2.6\times10^5$ entries, and---more
fundamentally---the classical model must represent the joint cortical state and its
couplings explicitly. This is precisely the regime that motivates a companion
quantum realisation of the same port-Hamiltonian model, in which the joint state is
carried in the amplitudes of $N$ qubits (a state space of dimension $2^N$) at a
parameter cost that stays linear in the functional network. The two models are the
same physics in two representations; the classical vector/matrix form developed here
is exact and sufficient at the montage scales this work evaluates, and is the
baseline against which any quantum benefit at whole-brain scale must be measured.

\section{Results}

We fit and score the twin entirely on real recordings: the
PhysioNet EEG Motor Movement/Imagery database
(EEGMMIDB;~\cite{schalk2004,goldberger2000}), $N=64$ channels at $160$\,Hz
(resampled to $f_s=250$\,Hz), $12$ subjects, $60$ recordings spanning eyes-open
rest, eyes-closed rest, and two motor-imagery paradigms. Every quantity reported
below is computed from recorded EEG: the canonical phasor state and the
phase-locking connectivity prior (this section), the model-independent dynamical
invariants that define the validation targets (Table~\ref{tab:real_targets}), and
the fitted-model scores against those targets (Section~\ref{sec:upgrade}). Splits
are leakage-free by subject: the held-out subjects used for testing never
contribute a sample to training.

\subsection{Dataset composition and phasor engineering}
\label{sec:dataset_overview}

\begin{figure}[htbp]
  \centering
  \includegraphics[width=\textwidth]{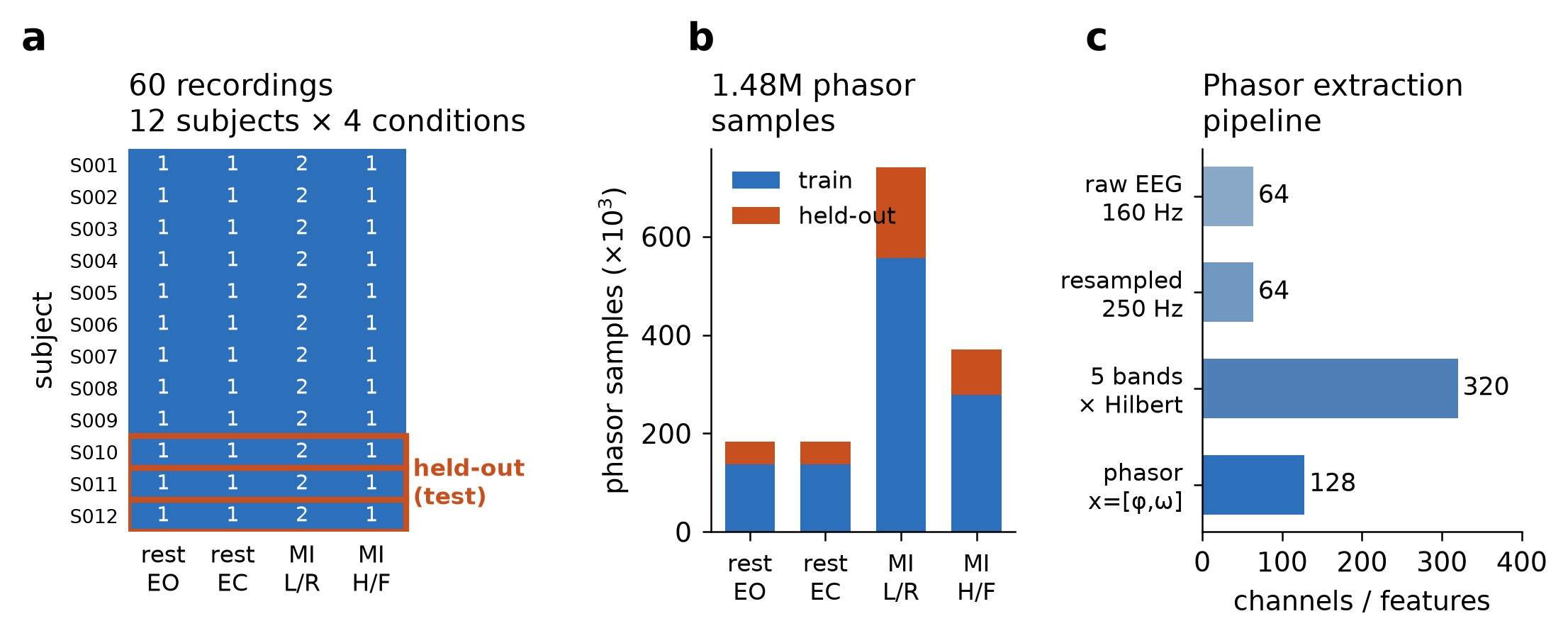}
  \caption{Dataset composition and phasor engineering (EEGMMIDB). \textbf{(a)}~The $60$-recording corpus as a subject~$\times$~condition grid (cell value = recordings); the three test subjects (\FitHoldout) are outlined and held out entirely. \textbf{(b)}~Phasor samples per condition after extraction, split into training and held-out test ($\approx1.48$\,M samples total). \textbf{(c)}~The preprocessing pipeline, tracing the feature dimension from the $64$-channel raw signal through the five-band Hilbert decomposition to the $128$-dimensional canonical phasor state $\mathbf{x}=[\bm{\phi},\bm{\omega}]$.}
  \label{fig:dataset_overview}
\end{figure}

Figure~\ref{fig:dataset_overview} summarises the data engineering that turns the
raw recordings into the design matrix. The corpus is a balanced
$12\times4$ block --- each of the $12$ subjects contributes one eyes-open-rest,
one eyes-closed-rest, and two motor-imagery recordings (left/right hand and
hands/feet), $60$ recordings in total (Fig.~\ref{fig:dataset_overview}a) --- and
the by-subject split holds subjects \FitHoldout\ out entirely, so no held-out
sample is seen in training (Fig.~\ref{fig:dataset_overview}b). Each recording is
average-referenced, band-pass filtered, resampled to $250$\,Hz, and separated
into the five canonical bands by Butterworth filters; a Hilbert transform then
yields the instantaneous phase and angular frequency that assemble the canonical
state $\mathbf{x}=[\bm{\phi},\bm{\omega}]\in\mathbb{R}^{128}$
(Fig.~\ref{fig:dataset_overview}c). The result is a single design matrix of
$\approx1.48$ million phasor samples, of which \FitTrainN\ are training and
$368{,}250$ are held-out test.

\subsection{Multi-band phasor extraction}

\begin{figure}[htbp]
  \centering
  \includegraphics[width=\textwidth]{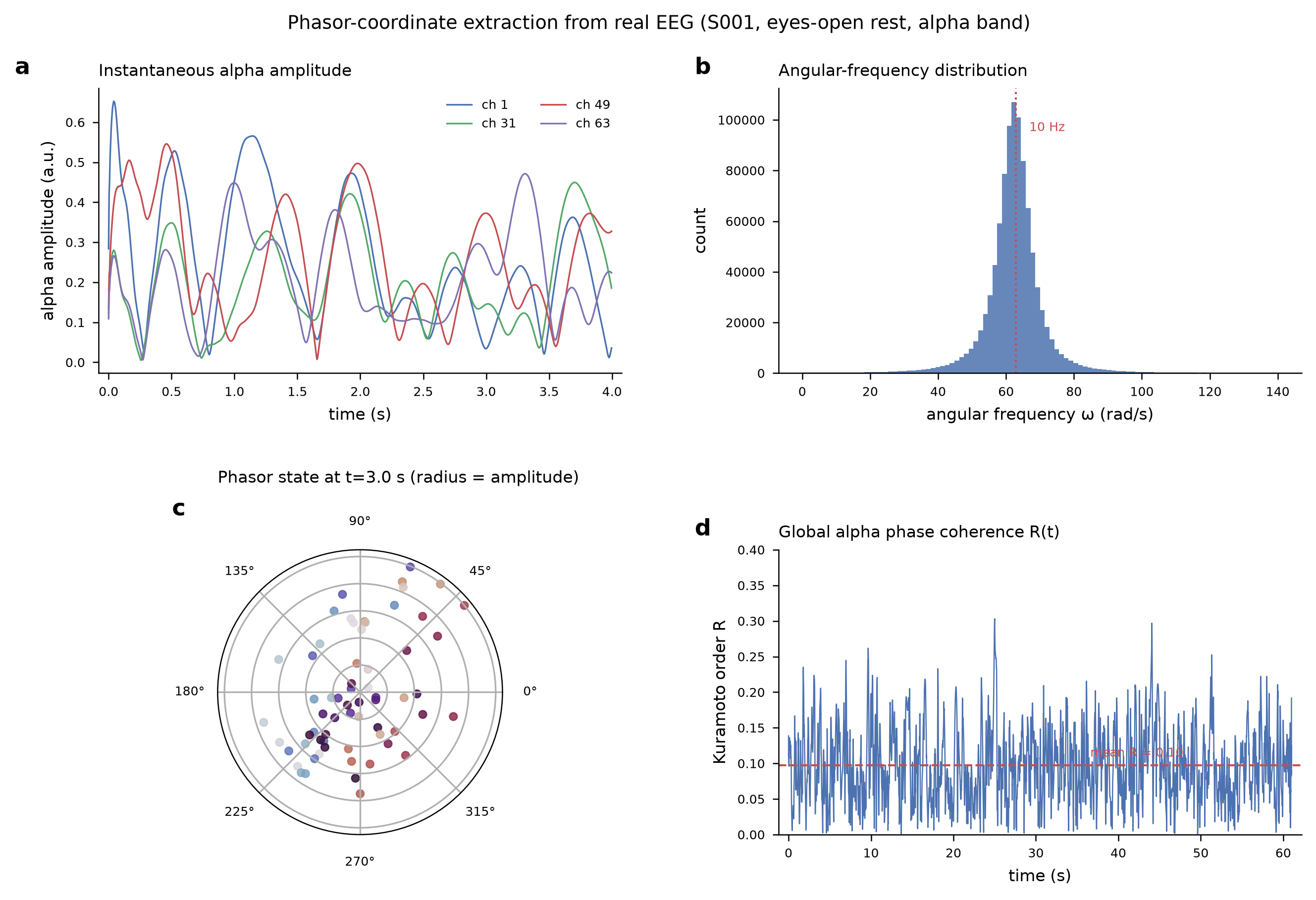}
  \caption{Phasor-coordinate extraction from EEGMMIDB (subject S001, eyes-open rest, alpha band). \textbf{(a)}~Instantaneous alpha amplitude for four channels. \textbf{(b)}~Angular-frequency distribution across all channels and time, peaking at the $10$\,Hz alpha centre (dotted line). \textbf{(c)}~Phasor state at $t=3.0$\,s: each channel is a point at angle $\phi$ and radius equal to its alpha amplitude. \textbf{(d)}~Global Kuramoto order parameter $R(t)$, fluctuating around $\bar R\approx0.10$ --- the partially synchronised waking regime. The phase and angular-frequency channels together are the canonical coordinates $\mathbf{x}=[\bm{\phi},\bm{\omega}]$.}
  \label{fig:phasor_extraction}
\end{figure}

Figure~\ref{fig:phasor_extraction} shows the phasor extraction from a
representative recording (subject S001, eyes-open rest). Each band-limited
analytic signal yields an instantaneous phase $\phi_j^{(b)}(t)$ and angular
frequency $\omega_j^{(b)}(t)$; together they form the canonical Hamiltonian
coordinates $\mathbf{x}=[\bm{\phi},\bm{\omega}]$. Two internal consistency checks
confirm a clean extraction: the alpha angular-frequency distribution peaks at
$2\pi\cdot 10\,\mathrm{rad\,s^{-1}}$ (the $10$\,Hz alpha centre), and the global
Kuramoto order parameter $R(t)$ fluctuates around a low mean
($\bar R \approx 0.10$), the partially synchronised regime characteristic of
awake cortex rather than the near-unity coherence of a pathological or degenerate
oscillator bank.

\subsection{PLV functional connectivity}

\begin{figure}[htbp]
  \centering
  \includegraphics[width=\textwidth]{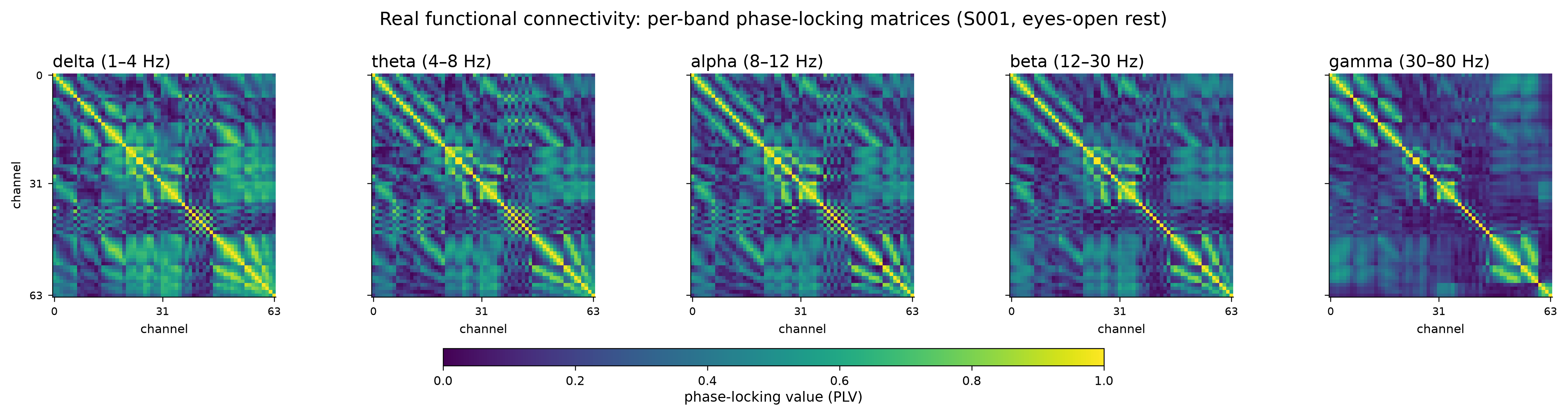}
  \caption{Functional connectivity from EEGMMIDB (subject S001, eyes-open rest): phase-locking-value matrices for the five canonical bands, sharing the colour scale (unit diagonal). Strong near-diagonal blocks (neighbouring electrodes are phase-locked) coexist with distributed off-diagonal coherent clusters (long-range functional coupling). These matrices provide the biologically principled coupling prior for the learned connectome $\mathbf{J}(\mathbf{x})$; the band-wise coherence gradient they encode is quantified in Fig.~\ref{fig:plv_gradient}.}
  \label{fig:plv}
\end{figure}

\begin{figure}[htbp]
  \centering
  \includegraphics[width=0.55\textwidth]{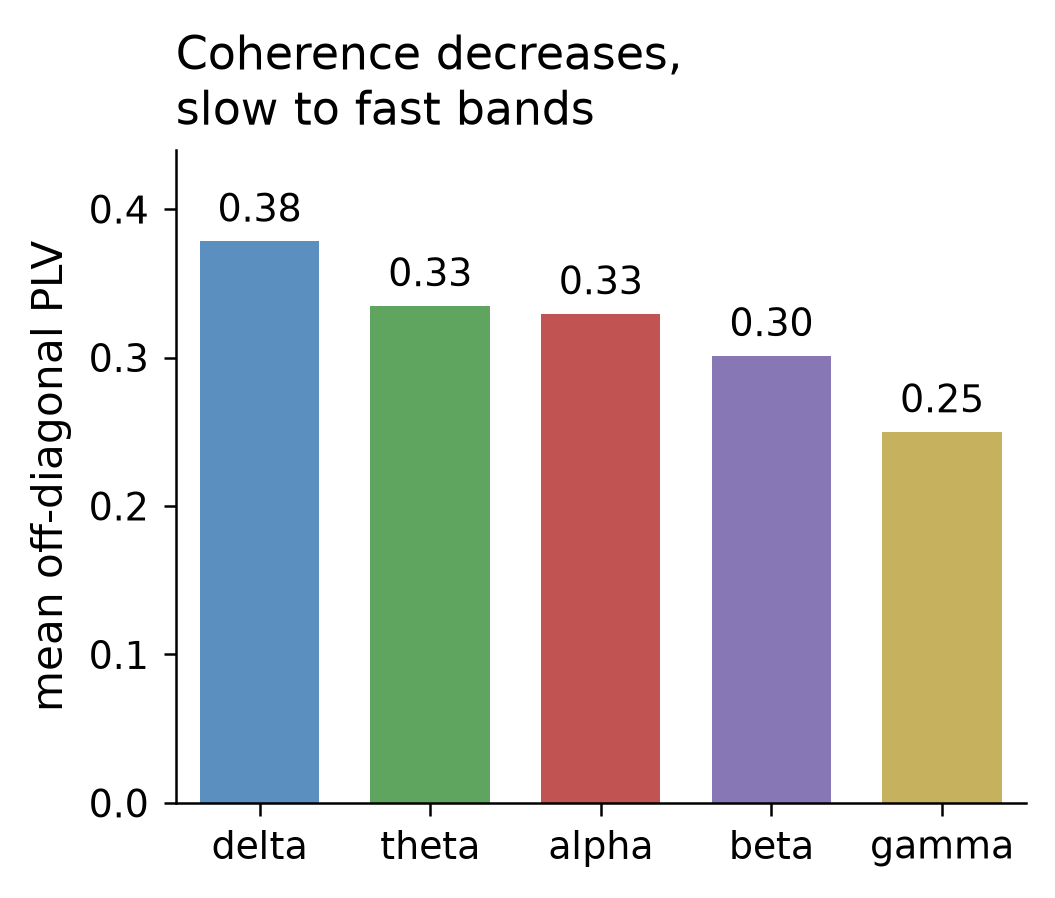}
  \caption{Mean off-diagonal PLV per band (EEGMMIDB, subject S001, eyes-open rest). Coherence decreases monotonically from delta through gamma --- delta $0.38 >$ theta $0.33 \geq$ alpha $0.33 >$ beta $0.30 >$ gamma $0.25$ --- the canonical spatial-scale gradient of cortical synchrony, in which low-frequency rhythms carry the most globally coherent, long-range synchrony while gamma coherence is spatially local.}
  \label{fig:plv_gradient}
\end{figure}

Figure~\ref{fig:plv} shows the empirical phase-locking-value (PLV) matrices for
the five bands, computed from EEGMMIDB. The matrices carry physiologically
structured connectivity: strong near-diagonal blocks (neighbouring electrodes are
phase-locked) together with distributed off-diagonal coherent clusters
(long-range functional coupling). The mean off-diagonal PLV falls monotonically
from the slow to the fast bands --- delta $0.38 >$ theta $0.33 \geq$ alpha $0.33 >$
beta $0.30 >$ gamma $0.25$ (Fig.~\ref{fig:plv_gradient}) --- reproducing the well-established result that
low-frequency rhythms carry the most globally coherent, long-range synchrony while
gamma coherence is spatially local. These matrices provide the coherence prior
that gates the dynamic connectome $\mathbf{J}(\mathbf{x})$.

\subsection{Dynamical invariants: the validation targets}
\label{sec:real_invariants}

\begin{figure}[htbp]
  \centering
  \includegraphics[width=\textwidth]{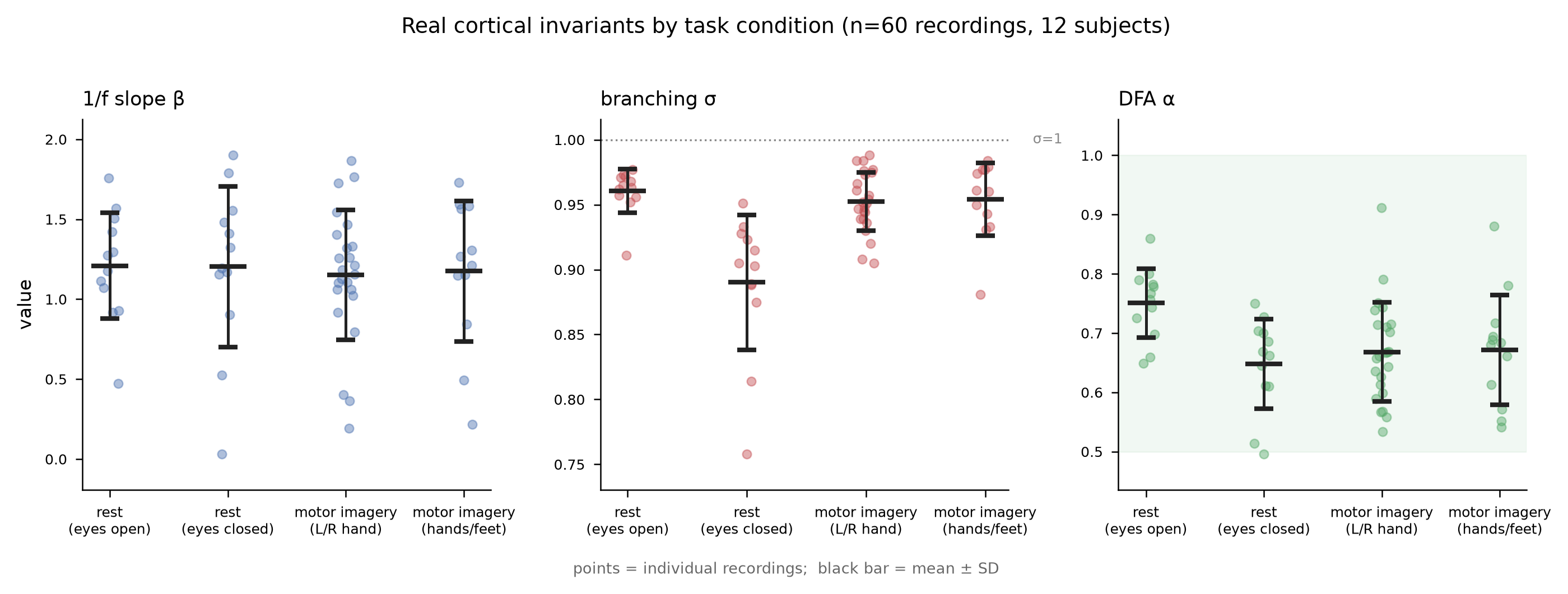}
  \caption{EEGMMIDB dynamical invariants across $60$ recordings ($12$ subjects), broken out by task condition. Left: aperiodic $1/f$ slope $\beta$. Middle: avalanche branching $\sigma$ (dotted line = criticality $\sigma=1$). Right: DFA exponent $\alpha$ (shaded long-range-temporal-correlation band $[0.5,1]$). Points are individual recordings; black bars are mean\,$\pm$\,SD. The cortex sits near criticality across all conditions; eyes-closed rest shows a modest reduction in branching and long-range temporal correlation relative to eyes-open rest.}
  \label{fig:real_invariants}
\end{figure}

The core of the analysis is to measure, on the recordings themselves, the
model-independent invariants that a twin claiming to track these data must
reproduce (Fig.~\ref{fig:real_invariants}, Table~\ref{tab:real_targets}). Across
all $60$ recordings the data sit close to the critical regime predicted by
the criticality hypothesis of cortical dynamics: the avalanche branching parameter
is $\sigma = 0.942 \pm 0.041$ (criticality at $\sigma = 1$), the detrended
fluctuation analysis exponent is $\alpha_{\mathrm{DFA}} = 0.68 \pm 0.09$ (within
the long-range-temporal-correlation band $[0.5,1]$), and the aperiodic spectral
exponent is $\beta = 1.18 \pm 0.42$ (the canonical cortical $1/f$ range). A
physiological task effect is visible: eyes-open rest carries the strongest
long-range temporal correlations (highest DFA), consistent with the
desynchronisation of temporal structure on eye closure. These three numbers ---
$\sigma$, $\alpha_{\mathrm{DFA}}$, and $\beta$ --- are the concrete rung-1 and
rung-4 targets the fitted twin is scored against in Section~\ref{sec:upgrade}.

\begin{table}[htbp]
\centering
\caption{Model-independent dynamical invariants measured on EEGMMIDB ($n=60$ recordings, $12$ subjects). These are the validation-ladder targets for the fitted twin.}
\label{tab:real_targets}
\begin{tabular}{llcc}
\toprule
\textbf{Rung} & \textbf{Invariant} & \textbf{Measured value} & \textbf{Critical/healthy target} \\
\midrule
1 & Aperiodic $1/f$ slope $\beta$      & $1.18 \pm 0.42$  & $\sim 1$--$1.5$ \\
4 & Avalanche branching $\sigma$        & $0.942 \pm 0.041$ & $\sigma \approx 1$ \\
4 & DFA exponent $\alpha_{\mathrm{DFA}}$ & $0.68 \pm 0.09$  & $0.5 < \alpha < 1$ \\
\bottomrule
\end{tabular}
\end{table}

\subsection{Fitting the metriplectic model to real phasors}
\label{sec:realfit}

\begin{figure}[htbp]
  \centering
  \includegraphics[width=0.72\textwidth]{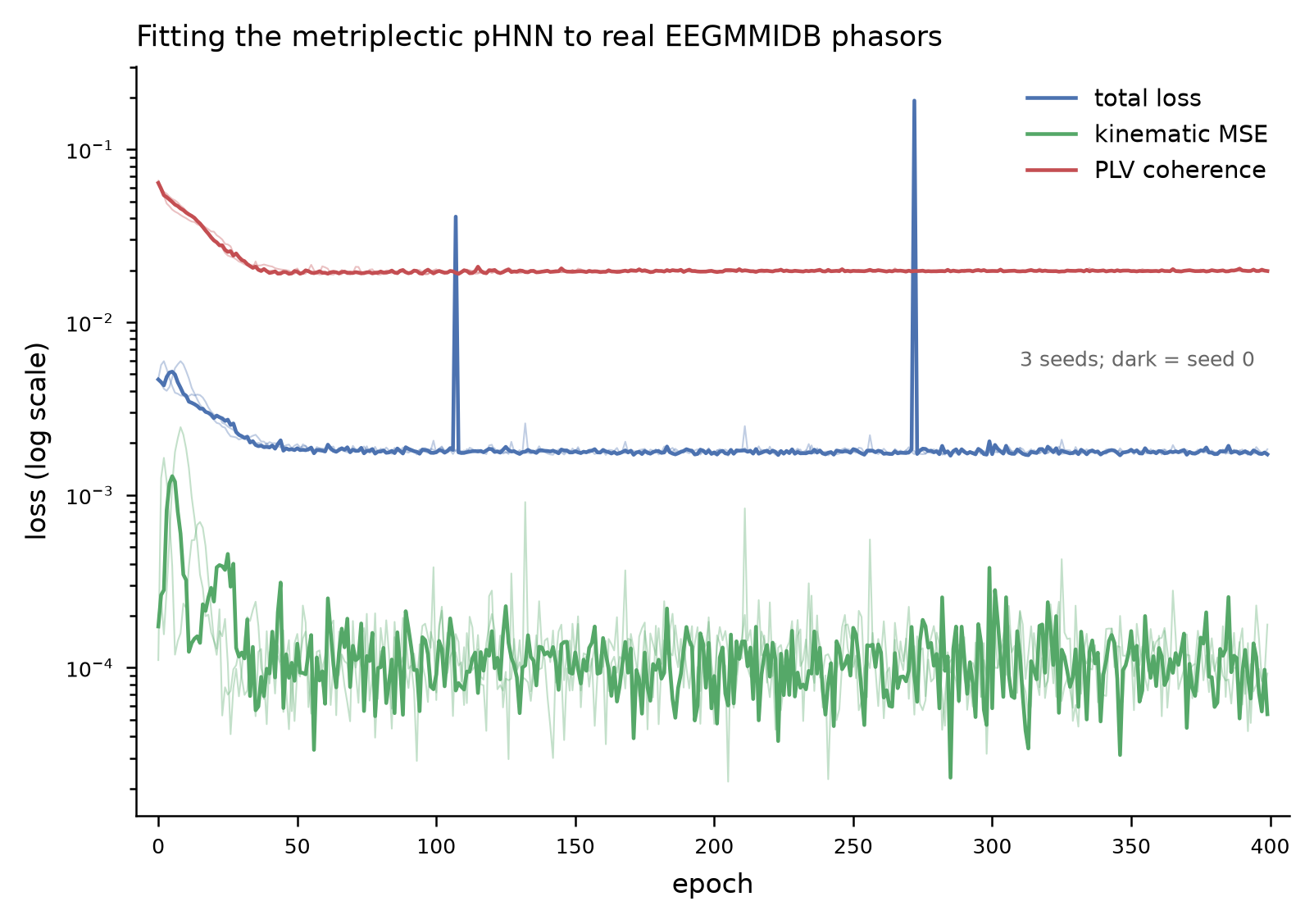}
  \caption{Training convergence fitting the metriplectic pHNN to EEGMMIDB phasors (logarithmic loss axis; three seeds, darker = seed 0). The composite objective and its kinematic, NESS-balance, and phase-coherence components are shown; all seeds converge to a common loss floor, with two brief transient excursions in seed 0 (near epochs $110$ and $270$) that recover within a few epochs. The held-out (by-subject) kinematic error is reported in the text.}
  \label{fig:training_loss}
\end{figure}

The metriplectic pHNN is fit to the canonical phasor state
$\mathbf{x}=[\bm{\phi},\bm{\omega}]$ under the composite physics objective of
Section~\ref{sec:methods} (kinematic reconstruction, the entropy-production /
NESS power balance, and the phase-coherence prior), with the reversible generator
constrained skew-symmetric and the dissipation positive-semidefinite by
construction. Training uses a leakage-free by-subject split
(\FitTrainN\ training samples; subjects \FitHoldout\ held out entirely, following
the protocol of Section~\ref{sec:fit_protocol}), and the
phase-coherence term $\mathcal{L}_\mathrm{PLV}$ is computed against the
\emph{measured} connectivity prior of Fig.~\ref{fig:plv} rather than a synthetic
one. Training and evaluation over the $\approx1.48$ million samples of the design
matrix are performed in mini-batches. The composite-loss trajectory and the held-out
(by-subject) kinematic error \FitTestMSE\ (mean over three seeds) are reported in
Fig.~\ref{fig:training_loss}; the three seeds converge to a common loss floor
(one seed shows two brief transient spikes that recover within a few epochs).

\begin{figure}[htbp]
  \centering
  \includegraphics[width=0.78\textwidth]{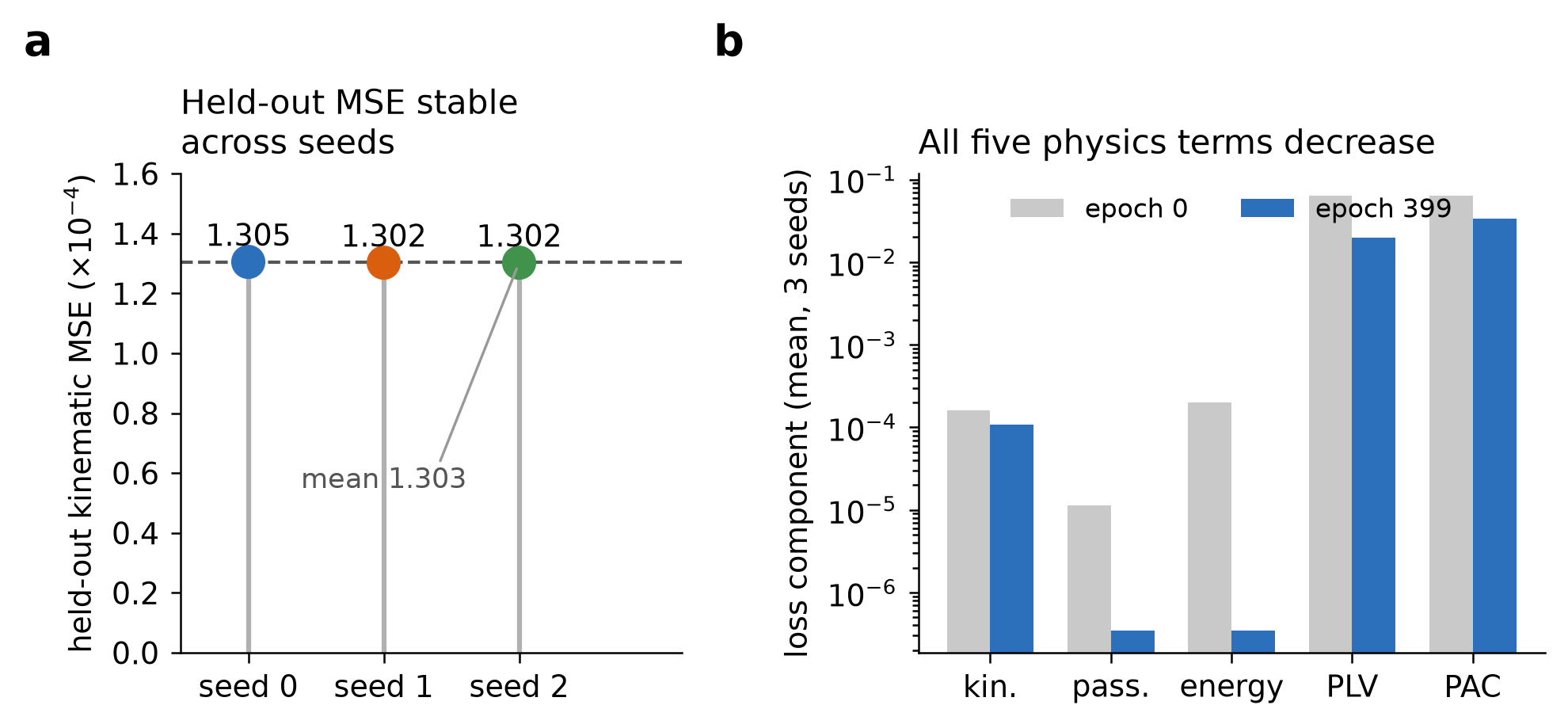}
  \caption{Seed robustness of the fit (three random seeds, $400$ epochs each; the composite-loss trajectories themselves are shown in Fig.~\ref{fig:training_loss}). \textbf{(a)}~Held-out (by-subject) kinematic MSE per seed, varying by under $0.3\%$ about the mean (dashed). \textbf{(b)}~The five physics-loss components (mean over seeds) at initialisation versus convergence; every term decreases.}
  \label{fig:training_robustness}
\end{figure}

The fit is robust to the random seed
(Fig.~\ref{fig:training_robustness}). The held-out kinematic MSE varies by under
$0.3\%$ across seeds ($1.305, 1.302, 1.302 \times10^{-4}$; mean \FitTestMSE,
Fig.~\ref{fig:training_robustness}a), and all five physics-loss terms ---
kinematic, passivity, energy balance, PLV coherence, and PAC --- decrease from
initialisation to convergence (Fig.~\ref{fig:training_robustness}b), confirming
that the composite objective is jointly minimised rather than trading one term
against another.

The remaining results --- reproduction of the $1/f$ spectrum, the learned
connectome against the measured PLV connectivity, and the criticality invariants
of the free-running model against the real targets of
Table~\ref{tab:real_targets} --- constitute the validation ladder and are reported
in Section~\ref{sec:upgrade}, where each rung is scored on the recordings.

\section{Validation on Cortical Dynamics and the Upgrade Path}
\label{sec:upgrade}

The twin of the preceding sections is fit to real recordings
(Section~\ref{sec:realfit}); what remains is to score it against
model-independent invariants that it did not author. We do this by
climbing a falsifiable validation ladder (Table~\ref{tab:validation_ladder}),
and we report below the rungs we can score on open EEG data --- spectral fidelity
(rung~1), functional connectivity (rung~2, via the measured PLV prior), and
criticality (rung~4) --- together with an honest account of where the fitted
twin succeeds and where it falls short.

It is worth restating what is being
certified: fitted to scalp EEG without a subject's own anatomical connectome, the
twin is not a replica of an individual brain, and the ladder does not test whether
it is one. What the ladder tests is whether a \textbf{physics-structured,
data-driven twin of the recordings} reproduces the dynamical invariants those
recordings carry --- a claim about fidelity to data, not about personal
identifiability or mechanism. The
subsections that follow the results set out the further latent-state,
connectivity, inference, and perturbational (rung~5, TMS-EEG) upgrades that open
data cannot yet close; each is accompanied by an explicit mathematical
specification establishing that the mechanism is well defined.

\subsection{Validation-ladder results}
\label{sec:ladder_results}

\begin{figure}[htbp]
  \centering
  \includegraphics[width=\textwidth]{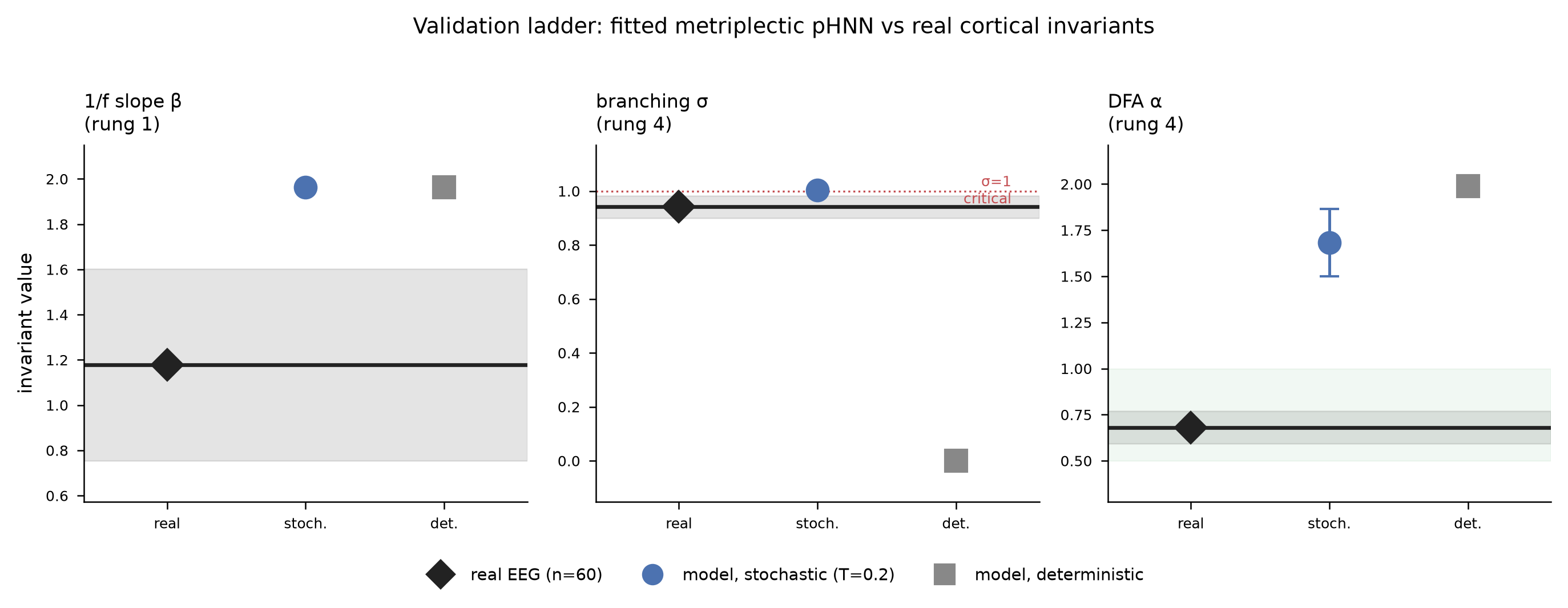}
  \caption{Validation ladder: the free-running fitted twin scored against invariants measured on the recordings (three seeds). For each rung the real target (diamond; shaded band $\pm1$\,SD over $60$ recordings) is compared to the stochastic free-run at $T=\FitTemp$ (circle) and the deterministic reference (square). The twin matches near-critical branching ($\sigma\approx1$) but produces too steep a $1/f$ slope and too high a DFA exponent.}
  \label{fig:validation_ladder}
\end{figure}

We score the \emph{free-running} fitted twin --- generating its own
trajectories from held-out initial conditions, with no teacher forcing ---
against the invariants measured on the recordings (Table~\ref{tab:real_targets},
Fig.~\ref{fig:validation_ladder}). Because criticality is a fluctuation
phenomenon, the theory-correct free-run of a metriplectic non-equilibrium
steady-state model is the fluctuation--dissipation-consistent stochastic rollout
(Eq.~\ref{eq:fdt}) at an arousal temperature $T=\FitTemp$; the noiseless
deterministic rollout is reported as a reference and, as expected, relaxes onto a
low-dimensional orbit that produces no avalanches.

The result is mixed and we state it plainly. \textbf{Criticality (rung~4,
branching): pass.} The stochastic free-run reproduces near-critical avalanche
dynamics, with a branching parameter $\sigma=\LadderSigmaModel$ against the real
$\LadderSigmaReal$ --- the metriplectic fluctuation structure generates
self-organised near-critical activity, not merely a fitted orbit.
\textbf{Spectral slope (rung~1): fail.} The model's aperiodic exponent
$\beta=\LadderOneModel$ is markedly steeper than the measured $\LadderOneReal$: the
twin's dynamics are spectrally too smooth, lacking the broadband high-frequency
power the recordings carry. \textbf{Long-range temporal correlations (rung~4, DFA):
fail.} The twin's DFA exponent $\alpha=\LadderDFAModel$ exceeds the measured
$\LadderDFAReal$ and lies above the stationary long-range-correlation band
$[0.5,1]$, indicating over-persistent, insufficiently itinerant dynamics. The
fitted twin therefore captures the \emph{branching} signature of criticality but
not the \emph{spectral} and \emph{temporal-correlation} signatures --- a concrete,
falsifiable gap that the excitation--inhibition and criticality-control upgrades
of Section~\ref{sec:criticality} are designed to close.

\begin{figure}[htbp]
  \centering
  \includegraphics[width=0.72\textwidth]{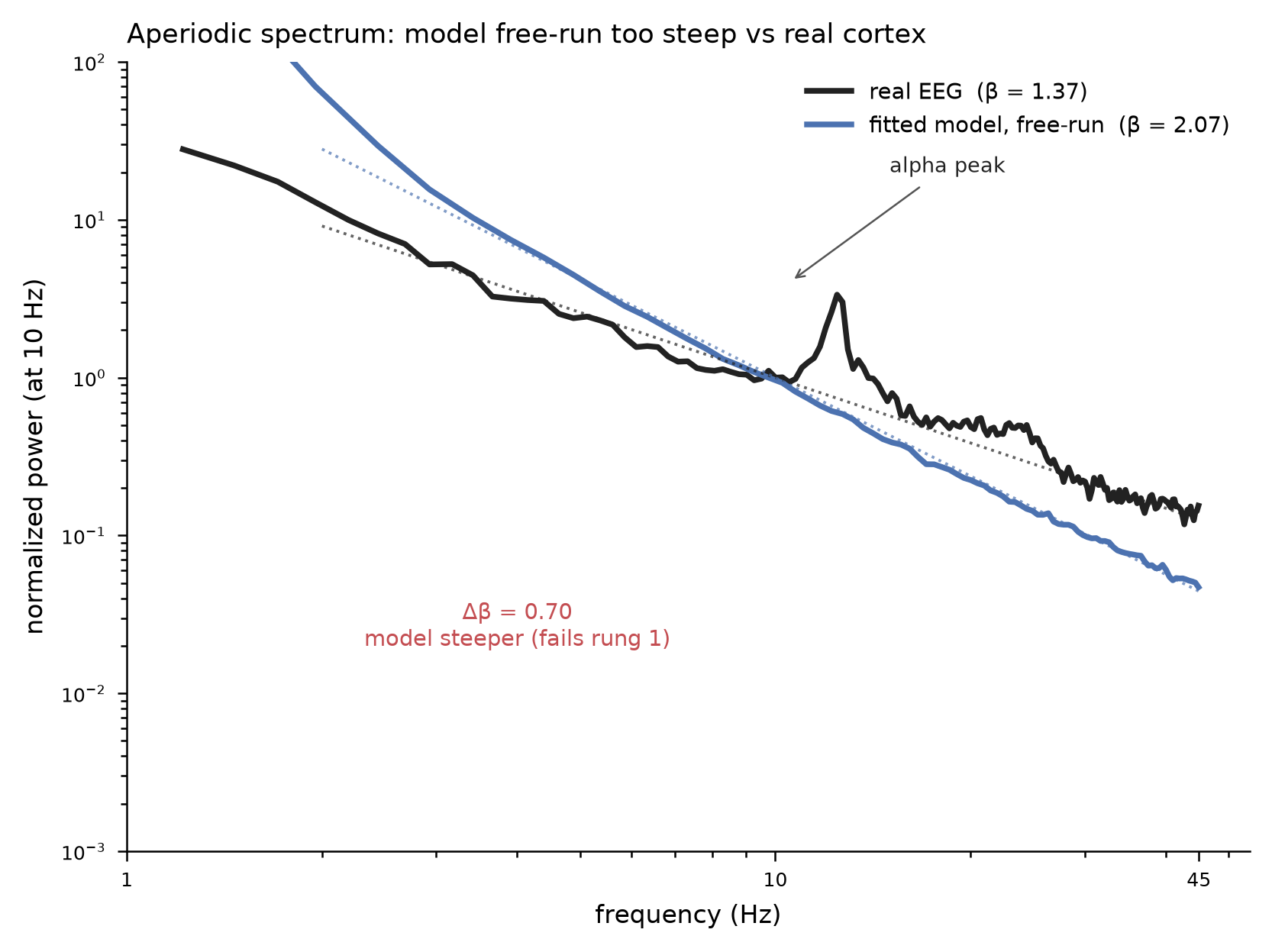}
  \caption{Aperiodic power spectrum, real EEG versus the fitted twin's stochastic free-run (both normalised to unit power at $10$\,Hz; dotted lines are the fitted $1/f^\beta$ slopes over $2$--$45$\,Hz). The real spectrum is shallower ($\beta\approx1.4$) and shows the characteristic alpha peak; the model spectrum is steeper ($\beta\approx2.1$) and lacks broadband high-frequency power. The slope gap $\Delta\beta$ is the rung-1 failure the criticality-control upgrade targets.}
  \label{fig:spectrum}
\end{figure}

The spectral failure is shown directly in Fig.~\ref{fig:spectrum}: the real EEG
power spectrum is comparatively shallow and carries a distinct alpha peak, whereas
the fitted twin's free-run spectrum is steeper and featureless. The two aperiodic
slopes differ by $\Delta\beta\approx0.7$. This is the concrete rung-1 gap; the
excitation--inhibition and criticality-control upgrades below (Section~\ref{sec:criticality})
are the mechanism designed to shallow the model spectrum toward the real one.

\begin{figure}[htbp]
  \centering
  \includegraphics[width=\textwidth]{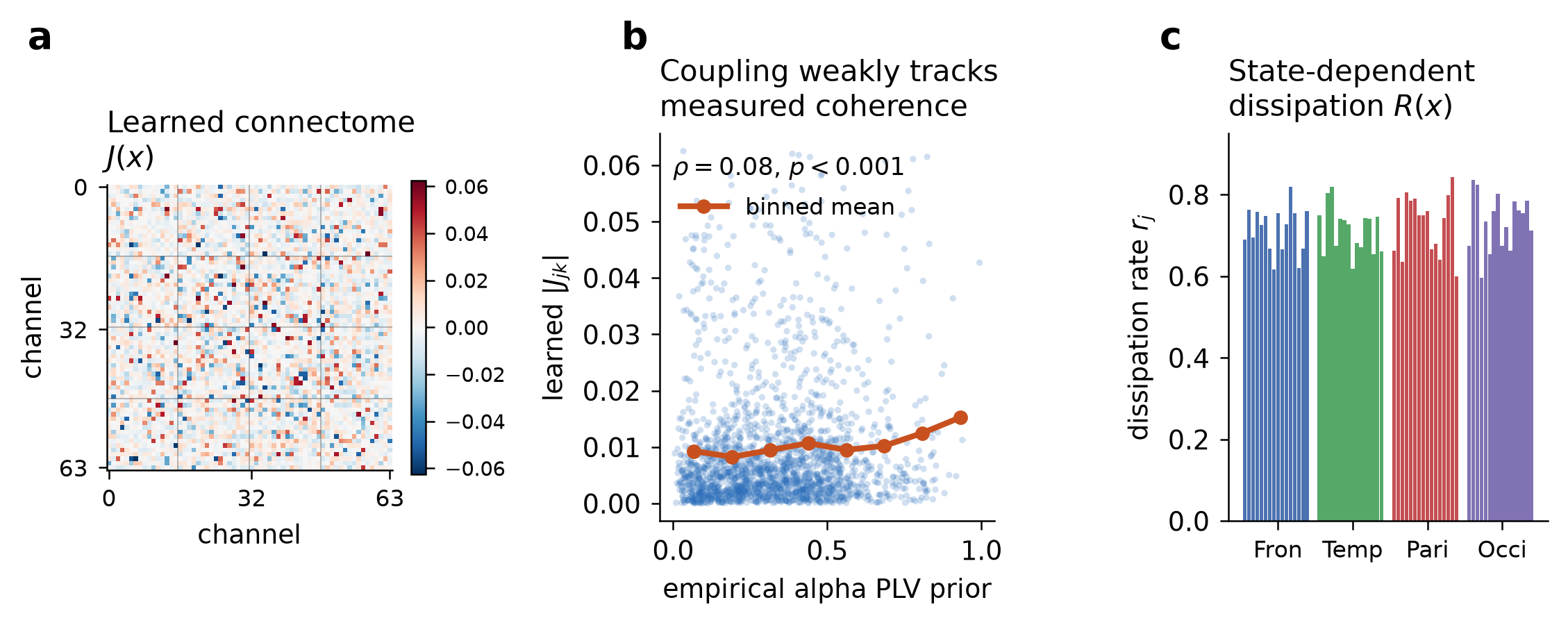}
  \caption{The learned operators of the fitted twin (seed~0). \textbf{(a)}~Learned functional connectome $\mathbf{J}(\mathbf{x})$ over the $64$ channels (angular-frequency coupling block), skew-symmetric by construction and gated by the measured alpha-band PLV; thin lines mark the frontal/temporal/parietal/occipital region boundaries. \textbf{(b)}~Learned coupling magnitude $|J_{jk}|$ versus the empirical alpha PLV prior for all channel pairs; the binned mean (orange) trends upward with the prior (Spearman $\rho=0.08$, $p<0.001$), showing that coupling is preferentially admitted where measured coherence is present. \textbf{(c)}~Per-channel dissipation rate $r_j$ grouped by scalp region, showing regionally heterogeneous, positive-semidefinite decay.}
  \label{fig:learned_connectome}
\end{figure}

Beyond the scored invariants, the fitted twin exposes its two learned physical
operators (Fig.~\ref{fig:learned_connectome}). The reversible generator
$\mathbf{J}(\mathbf{x})$ is skew-symmetric by construction
(Eq.~\ref{eq:skew_sym}) and, after the alpha-band PLV gate
(Eq.~\ref{eq:plv_gate}), carries a sparse, spatially structured coupling pattern
(Fig.~\ref{fig:learned_connectome}a). The gating leaves a measurable imprint: the
learned coupling magnitude $|J_{jk}|$ is weakly but significantly positively
associated with the empirical alpha PLV prior across all channel pairs
(Spearman $\rho=0.08$, $p<0.001$; binned mean rising from $\approx0.009$ at low
PLV to $\approx0.015$ at high PLV, Fig.~\ref{fig:learned_connectome}b) --- the
mechanism by which the model's functional connectome is anchored to measured
synchrony rather than free to invent coupling. The association is modest because
the gate suppresses, rather than prescribes, coupling: it removes support where
coherence is absent but does not force large coupling where it is present. The state-dependent dissipation $\mathbf{R}(\mathbf{x})\succeq0$
(Eq.~\ref{eq:r_net}) is regionally heterogeneous, with modestly higher decay over
parietal and occipital channels than frontal and temporal ones
(Fig.~\ref{fig:learned_connectome}c), consistent with the posterior localisation
of resting alpha dissipation.

\subsection{Reformulation: a non-equilibrium, stochastic model}

The single most consequential change is the move from strict global passivity to the non-equilibrium metriplectic formulation of Section~\ref{sec:metriplectic}, and three practical consequences follow directly. The first is autonomous sustainment rather than decay. A strictly passive autonomous system relaxes toward silence, whereas a twin that tracks resting recordings must instead sustain oscillation indefinitely. The diagnostic should accordingly be extended to a $30$--$60$\,s free rollout, and the reported invariant changed from $\max\dot{\mathcal{H}} \leq 0$ to orbital (Floquet) stability of the resting limit cycle together with the steady-state power balance of Eq.~\ref{eq:ness_balance}.

The second consequence is an explicit metabolic port: the metabolic influx $P_{\mathrm{met}}$ (Eq.~\ref{eq:ness_balance}) becomes a first-class model element rather than a residual, providing the anchor for calibrating $\mathcal{H}$ against haemodynamic and metabolic measurements such as fMRI BOLD, fNIRS, and cerebral metabolic rate of glucose --- the step that would convert the twin's Lyapunov storage function from an arbitrary scale into an interpretable energy budget. The third is thermodynamically consistent variability: the fluctuation--dissipation drive (Eq.~\ref{eq:fdt}) endows the model with trial-to-trial variability and metastable state switching \cite{Deco2011,Breakspear2017} at no additional parametric cost, with a single temperature $T$ standing in for arousal and neuromodulatory tone.

\subsection{A Latent Cortical State with an Explicit EEG Observation Model}
\label{sec:statespace}

The most consequential structural change concerns \emph{what the dynamics act upon}. Scalp EEG is a low-rank, instantaneously mixed, linear projection of a higher-dimensional cortical field, corrupted by sensor noise. Fitting the dynamics directly on the electrodes therefore embeds volume conduction, the common reference, and the lead field into the learned generator, so that the recovered connectome is partly an artefact of the montage rather than a property of the brain. The remedy is to separate the two explicitly: a latent cortical state $\mathbf{z}(t)$ carries the port-Hamiltonian / metriplectic dynamics of Section~\ref{sec:metriplectic}, and the EEG is an \emph{observation} of that state,
\begin{equation}
  \mathbf{y}(t) = \mathbf{C}\,\mathbf{z}(t) + \mathbf{e}(t),
  \label{eq:emission}
\end{equation}
where $\mathbf{C}$ is a lead-field-shaped observation operator and $\mathbf{e}$ is sensor noise.

Training then constrains the latent dynamics \emph{through} the emission, so the physics governs the cortical state that produces the EEG rather than the sensor traces themselves. An amortised encoder infers the initial latent state from a short EEG window, the metriplectic generator rolls it forward, and Eq.~\ref{eq:emission} emits predicted EEG at each step; the whole is trained by a prediction / evidence-lower-bound objective. This single move converts ``a model of EEG'' into ``a model of the latent cortical dynamics observed through EEG,'' and it is the natural home both for source localisation and for the fluctuation--dissipation noise of Eq.~\ref{eq:fdt}.

\subsection{Grounding the connectome in source-space measurements}
\label{sec:sourcespace}

\paragraph{Volume-conduction-robust coupling.} On scalp EEG, zero-lag phase locking is strongly inflated by volume conduction and the common reference, so the alpha-band PLV prior (Eq.~\ref{eq:plv}, Eq.~\ref{eq:plv_gate}) will, on scalp data, gate $\mathbf{J}(\mathbf{x})$ toward field spread rather than genuine coupling. The prior should be replaced by connectivity metrics that are insensitive to instantaneous mixing: the imaginary part of coherency \cite{Nolte2004}, the (weighted) phase-lag index \cite{Stam2007,Vinck2011}, or orthogonalised amplitude-envelope correlation. The phase-locking-gating mechanism itself is retained; only the estimator changes. The motivation is quantitative: in a controlled test, a pure zero-lag volume-conduction mixture yields a phase-locking value of $0.88$ but a weighted phase-lag index of only $0.015$, whereas a genuine $20$\,ms-lagged coupling yields a weighted phase-lag index of $0.999$ --- the replacement estimator rejects field spread while retaining true lagged coupling.

\paragraph{Source space and anatomical structure.} A stronger upgrade lifts the state out of sensor space entirely: inverse-solve to cortical sources (eLORETA / beamforming \cite{PascualMarqui2011}) on a parcellated cortex, so that nodes are anatomical regions rather than the four coarse electrode blocks of Eq.~\ref{eq:G_matrix}. The connectome $\mathbf{J}(\mathbf{x})$ then lives on the \textbf{structural connectome}: diffusion-MRI tractography supplies a fixed anatomical backbone that masks which entries \emph{may} be non-zero \cite{SanzLeon2013,Deco2013}, and fibre-length-derived \textbf{conduction delays} ($\sim$5--30\,ms) are introduced, moving the model into the delay-coupled neural-field regime \cite{JansenRit1995,WongWang2006} in which realistic inter-areal phase lags are achievable. This grounds the discovered functional connectome in anatomy rather than in synchrony statistics alone. Absent an individual's own tractography, the honest form of this constraint is a \emph{population-level} one: a group-average structural connectome supplies a soft prior on which entries of the coupling $\mathbf{J}(\mathbf{x})$ may be large, so that the model respects the typical anatomical backbone without claiming a subject-specific replica. The associated penalty vanishes for coupling confined to the backbone and is large for anatomically unsupported coupling. Inter-node distances on a spherical cortical layout, divided by a conduction velocity of $\sim$6\,m/s, produce conduction delays spanning the physiological $0$--$30$\,ms range (mean near $17$\,ms) that enter the reversible generator $\mathbf{J}(\mathbf{x})$, placing the model in the delay-coupled neural-field regime.

\subsection{Excitation--inhibition balance and the critical operating point}
\label{sec:criticality}

The property that most distinguishes the recordings from a curve-fit is that they are not a fixed limit cycle but an itinerant trajectory poised at the edge of a phase transition: neuronal avalanches with a branching parameter $\sigma\!\approx\!1$, long-range temporal correlations, and $1/f$-like spectra \cite{BeggsPlenz2003,Linkenkaer2001}. Two additions bring this within the twin's reach. First, each node is resolved into \textbf{excitatory and inhibitory compartments} with structured Wilson--Cowan-type coupling; unlike the frequency-band decomposition, which is a filter-bank view of a single field, the excitation--inhibition split corresponds to genuinely distinct neural populations, making it the better-motivated composite decomposition of the twin's energy, $\mathcal{H}=\mathcal{H}_E+\mathcal{H}_I+\mathcal{H}_{EI}$.

A per-node operating-point parameter sets the balance, and hence the gain, and hence the proximity to the critical manifold. Second, a control objective actively \emph{tunes} that operating point toward criticality, rather than merely measuring criticality after the fact, together with differentiable regularisers that penalise deviation of the branching parameter and the aperiodic slope from their healthy-cortex values. The differentiable spectral-slope proxy that anchors this regulariser tracks the model-independent estimate to within numerical tolerance, so the training signal is faithful to the quantity it controls. This is the mechanism directly targeting the rung-1 spectral gap (Fig.~\ref{fig:spectrum}) and the rung-4 DFA gap of the current fit.

\subsection{Slow arousal and the timescale hierarchy}
\label{sec:slow}

The single temperature $T$ of the fluctuation--dissipation drive (Eq.~\ref{eq:fdt}) is, physically, a slow dynamical variable: ascending neuromodulatory tone on a timescale of seconds to minutes, and, over long recordings, a circadian rhythm. Promoting it from a fixed scalar to a \textbf{slow compartment} that modulates the fast dynamics' operating point places distinct recording conditions --- eyes-open versus eyes-closed rest, quiescence versus motor engagement --- as locations on a continuous arousal axis, and supplies the genuine slow ``clock'' of the brain. The natural realisation is not a set of autonomous separable oscillators --- cortical rhythms are emergent and strongly coupled --- but nested cross-frequency coupling, in which a slow phase gates a faster amplitude in a delta--theta--gamma cascade.

\subsection{Subject-specific inference}

The twin presented here is fitted across subjects; a \emph{subject-specific} twin adds a requirement of identifiability: given one individual's resting EEG, it must recover \emph{that person's} parameters --- individual alpha frequency, individual structural connectome, individual excitation/inhibition balance --- with calibrated uncertainty. This is an amortised Bayesian inverse problem. Simulation-based inference \cite{Cranmer2020} trains a posterior estimator over the physical parameters $(\mathbf{J}_0, \mathbf{R}_0, \text{delays}, T)$ from simulated EEG summary statistics, after which per-subject fitting is a single forward pass. Point estimates are insufficient for a clinical twin: therapeutic decisions require the posterior, so that a proposed stimulation is accompanied by its predictive uncertainty.

\subsection{A Falsifiable Validation Ladder}

Fidelity to the data is certified by climbing a sequence of increasingly stringent, model-independent targets (Table~\ref{tab:validation_ladder}). Each rung is a property the twin did not author, and each has an established public-data benchmark. The ladder is what keeps ``digital twin'' from being a slogan: it converts the word into a list of measurements the twin can fail.

\begin{table}[H]
  \centering
  \small
  \begin{tabular}{p{0.5cm} p{3.4cm} p{7.5cm}}
  \toprule
  \# & Target & Falsifiable criterion \\
  \midrule
  1 & Spectral fidelity & Reproduce the real power spectrum: $1/f$ aperiodic slope \emph{and} band peaks, not idealised rhythms \cite{Donoghue2020}. \\
  2 & Connectivity      & Recover source-space functional connectivity \emph{and its state modulation} on held-out subjects. \\
  3 & Forecasting       & Non-trivial multi-step prediction horizon on unseen EEG. \\
  4 & Criticality       & Reproduce neuronal avalanches and long-range temporal correlations (DFA exponents, power-law scaling) \cite{BeggsPlenz2003,Linkenkaer2001}. \\
  5 & Causal perturbation & Predict the TMS-evoked potential and Perturbational Complexity Index (PCI) under stimulation \cite{Casali2013}. \\
  6 & Generalisation    & Cross-subject / cross-session transfer without refitting the architecture. \\
  \bottomrule
  \end{tabular}
  \caption{Validation ladder defining what it takes for the twin to be called faithful to the data. Rung~5 (TMS-EEG) is the decisive test of the neuromodulation ports and the dissipative structure under perturbation, and is the only rung that directly validates the stimulation-simulation claim.}
  \label{tab:validation_ladder}
\end{table}

The diagnostics that score rungs~1, 3, and 4 are the same estimators applied to the fitted twin in Section~\ref{sec:ladder_results}: the aperiodic-slope fit, the avalanche branching parameter, and detrended fluctuation analysis, together with a long-horizon free rollout that must sustain bounded oscillation rather than decay to silence.

Rung~4 warrants elevation from a validation target to a modelling goal in its own right, because reproducing the \emph{dynamical repertoire} --- not a single resting orbit --- is the sharper test. A twin that tracks the data must wander through a repertoire of transiently visited, marginally stable states, yielding a broad dwell-time distribution and a structured functional-connectivity-dynamics matrix rather than a static connectome. The metastability and criticality objectives of Section~\ref{sec:criticality} are precisely the terms that detect an absent repertoire and supply the gradient toward it; they are the natural next addition given that the current fit clears branching but not the DFA and spectral rungs.

Rung~5 deserves emphasis. The abstract claims a substrate for closed-loop stimulation, yet no observational recording --- however real --- tests the ports $\mathbf{G}$ under an actively delivered perturbation. Concurrent TMS-EEG provides exactly this: the model must predict the spatiotemporal evoked response and its complexity to a known cortical perturbation. This is the single most convincing evidence that $\mathbf{G}$ and the metriplectic structure are physical rather than decorative.

\subsection{Control: Energy-Shaping for Closed-Loop Neuromodulation}

A therapeutic arousal-regulation objective --- steering a drowsy or dysregulated cortex back toward a healthy resting operating point --- is naturally posed as \textbf{energy-shaping / Interconnection-and-Damping-Assignment Passivity-Based Control} (IDA-PBC) \cite{Ortega2002}, the control theory native to port-Hamiltonian systems. Rather than a black-box inverse, IDA-PBC synthesises a stimulation $\mathbf{u}^*(t)$ that reshapes the closed-loop energy landscape so the desired resting limit cycle becomes the stable attractor, \emph{with stability guarantees}. Retaining the pH/metriplectic structure (as opposed to a generic recurrent surrogate) is precisely what makes these guarantees available, and the port matrix $\mathbf{G}$ should be upgraded from anatomical averaging (Eq.~\ref{eq:G_matrix}) to a physically modelled lead field --- the head-model-projected electric field for tDCS/TMS/DBS \cite{Thielscher2015}. A structural controllability analysis makes the binding constraint explicit: with the three anatomical ports of Eq.~\ref{eq:G_matrix}, only about $5\%$ of the $64$-node state space is directly reachable (reachability residual near $0.97$). This quantifies why the lead-field refinement and additional ports are prerequisites for the causal-control claim of rung~5, and why a closed-loop convergence result is deferred until after the source-space fit.

\subsection{Surrogate upgrades for scalability and speed}

Three changes make the GNN surrogate fit for cross-subject, faster-than-real-time closed-loop use. (i) \textbf{Geometric equivariance}: the connectome has no canonical node ordering, so an equivariant graph network that encodes electrode/parcel geometry is preferable to order-dependent attention. (ii) \textbf{Neural-operator formulation}: learning the \emph{solution operator} rather than the per-step vector field \cite{Li2021} enables amortised generalisation across subjects and horizons and supports real-time control. (iii) \textbf{Multi-rate integration}: the $\gamma$ and $\delta$ bands span two decades of timescale, so the stiff dynamics require multi-rate or exponential integrators for stable long-horizon rollouts.

\subsection{Summary of the Upgrade Path}

In priority order: (1)~adopt the latent state-space formulation with an explicit EEG emission operator (Section~\ref{sec:statespace}), so the physics governs the cortical state rather than the sensor traces, together with the metriplectic / NESS reformulation and its metabolic port; (2)~resolve excitatory and inhibitory compartments and add the criticality control that tunes the operating point toward the critical manifold (Section~\ref{sec:criticality}), which is the mechanism for the most important missing property; (3)~replace PLV with a volume-conduction-robust estimator and impose the population-level structural-connectome prior with conduction delays; (4)~fit to a public dataset (e.g.\ BCI Competition~IV \cite{Blankertz2007} or a resting/N-back corpus) and clear rungs~1--4 of Table~\ref{tab:validation_ladder}, including the metastability and dwell-time targets; (5)~add fluctuation--dissipation stochasticity with the slow arousal compartment (Section~\ref{sec:slow}) and subject-specific posterior inference; (6)~attempt the TMS-EEG causal validation (rung~5).

\section{Discussion}
\label{sec:discussion}

\subsection{What a Structural Guarantee Buys That a Penalty Does Not}

Figure~\ref{fig:architecture} draws a horizontal rule through the construction,
and that rule is the organising distinction of this work. Above it, two
properties are carried by the parameterisation. The antisymmetrisation
$\mathbf{J}=\mathbf{U}-\mathbf{U}^\top$ (Eq.~\ref{eq:skew_sym}) is skew for
every $\mathbf{U}$, and the softplus of Eq.~\ref{eq:r_net} is non-negative for
every input, so $\mathbf{R}(\mathbf{x})\succeq0$ at every state. A weight
configuration that violates either has no preimage in the parameter space. The
guarantee therefore holds at initialisation, after every gradient step, at every
state the twin visits during a free rollout, and at operating points arbitrarily
far outside the recordings it was fitted to. Below the rule sit the five terms of
Eq.~\ref{eq:composite_loss} --- kinematic accuracy, passivity, the energy
balance, and the two coherence priors. Each of these is a price the optimiser can
pay or decline. None is guaranteed anywhere, and each holds only over the region
of state space that the training data actually constrained.

The practical consequences separate along three lines. A structural property
needs no verification: it is audited by reading the construction, not by running
an experiment at every distribution of interest. It carries no distribution with
it, so it survives a shift of montage, condition, or subject. And there is no
constraint weight attached to it, hence no regime in which the physics is quietly
traded against the kinematic fit. A penalised property has none of these
attributes. Its residual is a measurement, valid where it was measured, and its
weight ($\lambda_p\!=\!0.5$, $\lambda_\mathrm{eb}\!=\!0.1$,
$\lambda_\mathrm{PLV}\!=\!0.05$, $\lambda_\mathrm{PAC}\!=\!0.02$) fixes the
exchange rate at which the fit may buy it back. This is why we report the
penalised half rather than assuming it: Fig.~\ref{fig:training_robustness}b shows
that all five terms decrease from initialisation to convergence, which
establishes that the objective is jointly minimised, and establishes nothing
beyond that.

Passivity is the instructive case, and the manuscript's own position on it should
not be overstated. The passivity term $\mathcal{L}_p$ penalises positive
$\dot{\mathcal{H}}$ at zero stimulation, and it decreases under training; we do
not report a certified bound $\dot{\mathcal{H}}\leq0$ over the reachable set,
because no penalty can supply one. More importantly, strict global passivity is
not the invariant a twin of waking recordings should want
(Section~\ref{sec:metriplectic}): it is the condition for relaxation to a fixed
point, and the deterministic free-run of Section~\ref{sec:ladder_results} does
relax onto a low-dimensional orbit that produces no avalanches at all. The
near-critical branching we do report comes from the
fluctuation--dissipation-consistent stochastic rollout, not from the passive
limit. A strong passivity measurement would therefore be evidence against the
model being what we claim, which is precisely why the operative invariant is the
steady-state power balance of Eq.~\ref{eq:ness_balance} rather than monotone
energy decay.

The guarantees also cost something, and the cost is expressiveness. A skew
$\mathbf{J}$ cannot represent a reversible generator with a symmetric part, and a
diagonal $\mathbf{R}\succeq0$ cannot represent negative local damping. The
multiplicative gate of Eq.~\ref{eq:plv_gate} is a third commitment of the same
kind: coupling is admissible only on the support of the prior, so an entry the
prior zeroes cannot be recovered by the fit however strongly the data ask for it.
These are modelling commitments, and stating them here is preferable to
discovering them later.

\subsection{$\mathbf{J}$ as an Internal Anatomical Connectome}
\label{sec:internal_connectome}

The gate that makes $\mathbf{J}(\mathbf{x})$ sparse is currently supplied by the
alpha-band phase-locking matrix (Eq.~\ref{eq:plv_gate}), and that choice sets a
ceiling on what the learned operator can mean. The phase-locking value is a
\emph{functional} quantity, measured at the scalp, and zero-lag coherence between
two electrodes is inflated wherever both see the same source: high coherence is
then evidence of shared field spread rather than of coupling between two cortical
regions \cite{Nolte2004,Stam2007}. Because the gate multiplies, this confound
does not merely add noise to $\mathbf{J}$ --- it decides which entries of
$\mathbf{J}$ are permitted to be non-zero at all. The support of the twin's
connectome is therefore, in part, a property of the montage.

The alternative is to let an \emph{internal anatomical} connectome supply the
mask instead: a diffusion-MRI tractography matrix, or a cortical parcellation
whose admissible edges are the known white-matter pathways between parcels
\cite{SanzLeon2013,Deco2013}. The reason this is a credible next step rather than
a speculative one is structural. It is a change of \emph{values}, not of
\emph{shapes}. The mask enters at one call site, Eq.~\ref{eq:plv_gate}, as an
$N\times N$ matrix multiplying $\mathbf{J}$ elementwise; a tractography matrix
has the same dimensions, the same range after normalisation, and the same
sparsity role. The operator dimensions, the energy network, the dissipation
model, the port matrix, and the loss all remain as written. Only the matrix that
enters the gate differs.

Two things would change if it did. First, $\mathbf{J}$ would acquire an
interpretation it does not currently have: routing of stored energy along
anatomical pathways, rather than admissible coupling wherever scalp signals were
co-observed. Second, because a parcellation and its pathways are defined
independently of any one recording, the twin's couplings would become comparable
across subjects --- the same matrix entry would denote the same anatomical
connection in every fit, which is not true of a montage-indexed phase-locking
matrix.

One thing would not change, and it is the more consequential half. Replacing the
mask does not solve source localisation. The state $\mathbf{x}=[\bm{\phi},
\bm{\omega}]$ is still assembled from electrode signals, so a mask defined on
cortical parcels and a state defined on scalp channels do not live on the same
index set. Making them agree requires the observation model of
Eq.~\ref{eq:emission} --- a lead-field-shaped operator mapping cortical sources
to electrodes --- which is a separate and additional requirement, not a
by-product of changing the gate. The upgrade path of
Section~\ref{sec:sourcespace} already specifies the source-space and structural
connectivity route in detail, including the population-level form appropriate
when an individual's own tractography is unavailable, and we do not restate it
here. The point of the present subsection is narrower: the interface through
which such a connectome would enter already exists, and it is one matrix wide.

\subsection{Scale: What the Twin Becomes at $10^3$ to $10^6$ Channels}
\label{sec:scale}

It is worth stating what this model becomes at whole-brain channel counts,
because the answer is not the one the $\mathcal{O}(N^2)$ figure of
Section~\ref{sec:complexity} suggests. Table~\ref{tab:scaling} tabulates the
objects from the $N{=}64$ montage evaluated here to a hypothetical
$N{=}10^{6}$-channel recording. Three features of that table matter, and the
third is a condition on the first two.

\begin{table}[htbp]
\centering
\caption{Scaling of the twin's objects with channel count $N$. Dense-$\mathbf{J}$
parameters are the upper-triangle entries of the $2N\times2N$ operator,
$N(2N-1)$. Masked parameters are the $|E|$ entries surviving the gate, at fixed
mean degree $20$, so $|E|=10N$. Dense operator storage is $(2N)^2$ single-precision
entries. The $N{=}64$ row is the configuration fitted in this work; the remaining
rows are projections under the stated assumption, not measurements.}
\label{tab:scaling}
\setlength{\tabcolsep}{8pt}
\begin{tabular}{rcrrrr}
\toprule
$N$ & State & Dense-$\mathbf{J}$ params & Masked $|E|$ & Reduction & Dense storage \\
\midrule
$64$        & $\mathbb{R}^{128}$          & $8.13\times10^{3}$ & $640$              & $12.7\times$            & $6.6\times10^{-5}$\,GB \\
$256$       & $\mathbb{R}^{512}$          & $1.31\times10^{5}$ & $2.56\times10^{3}$ & $51.1\times$            & $1.1\times10^{-3}$\,GB \\
$1{,}024$   & $\mathbb{R}^{2{,}048}$      & $2.10\times10^{6}$ & $1.02\times10^{4}$ & $205\times$             & $1.7\times10^{-2}$\,GB \\
$10^{4}$    & $\mathbb{R}^{20{,}000}$     & $2.0\times10^{8}$  & $1.0\times10^{5}$  & $2.0\times10^{3}\times$ & $1.6$\,GB \\
$10^{6}$    & $\mathbb{R}^{2\times10^{6}}$ & $2.0\times10^{12}$ & $1.0\times10^{7}$  & $2.0\times10^{5}\times$ & $1.6\times10^{4}$\,GB \\
\bottomrule
\end{tabular}
\end{table}

First, growth falls entirely on one component. The energy network is
size-invariant: its attention weights are shared across nodes
(Eq.~\ref{eq:self_attn}), so it stays at order $1.2\times10^{6}$ parameters
whether $N$ is $64$ or $10^{6}$. The dissipation rates, the three-column port
matrix and the gate scale as $\mathcal{O}(N+|E|)$ and remain negligible. What
grows is the predictor that emits $\mathbf{J}$, and how fast it grows is a
choice.

Second, dense versus masked is the whole story. A predictor that emits every
upper-triangle entry of the $2N\times2N$ operator carries $N(2N-1)$ parameters
--- $2.0\times10^{12}$ at a million channels, some $8$\,TB at single precision,
which is infeasible on any foreseeable device. A predictor that emits only the
$|E|$ entries surviving the gate carries $1.0\times10^{7}$ for the same operator,
so the whole model is about $1.1\times10^{7}$ parameters, roughly $45$\,MB at
single precision. The reduction is a factor of $2.0\times10^{5}$, and it costs
nothing, because Eq.~\ref{eq:plv_gate} sets the entries the predictor declines to
emit to zero in any case. Note that the $\mathcal{O}(N^{2})$ figure of
Section~\ref{sec:complexity} is dense-operator \emph{storage}, a memory cost, and
the last column of Table~\ref{tab:scaling} is that quantity; the parameter
columns are a different accounting and the two should not be conflated.

Third, and this is the assumption everything above rests on: the masked figures
hold mean degree \emph{fixed} at approximately $20$, the small-world regime, so
that $|E|$ grows linearly in $N$. That assumption is what makes the reduction
factor grow with $N$ rather than saturate. If dense high-channel recording were
to reveal genuinely dense coupling at fine spatial scales --- mean degree growing
with $N$ rather than staying constant --- then $|E|$ would grow as $N^{2}$, the
masked and dense columns would converge, and the favourable scaling would fail
outright. The favourable arithmetic is a consequence of a sparsity hypothesis
about cortical coupling, not a property of the architecture, and it is falsifiable
by exactly the recordings that would motivate the extrapolation.

Identifiability runs opposite to intuition, and it is worth stating why. Counting
observed scalars as $2N$ per sample over a corpus of fixed duration
($T\approx1.48\times10^{6}$ samples, as here), and masked parameters as the
size-invariant core plus $|E|$, the free parameters per observed scalar fall from
about $6\times10^{-3}$ at $N{=}64$ to about $4\times10^{-6}$ at $N{=}10^{6}$ --- a
drop of some three orders of magnitude. The mechanism is that observations grow
as $N\times T\times$~recordings while masked parameters grow as $N$ with a fixed
core amortised over more nodes. More channels is therefore \emph{more}
constraining, not less, and the configuration fitted in this work is the most
over-parameterised point on that curve. Two caveats bound the claim. It holds
only while the sparsity hypothesis above holds. And it is a counting argument
about degrees of freedom, not a guarantee of practical identifiability: nothing in
it prevents the parameters from being poorly conditioned, nor the additional
channels from being spatially redundant.

One implication for the companion quantum realisation follows.
Section~\ref{sec:complexity} motivates that realisation by the cost of
representing the joint state and its couplings explicitly, and the last column of
Table~\ref{tab:scaling} confirms that the \emph{dense} operator becomes
prohibitive. Under the masked parameterisation, however, the classical model
remains within a single device well beyond $10^{4}$ channels. The case for a
quantum representation should therefore rest on the joint state and its
correlations rather than on classical operator storage.

\subsection{What the Present Verdict Does and Does Not Support}

The evidence reported here separates into claims of different strength, and
conflating them would misrepresent the twin. The structural properties hold by
construction and need no data: exact skew-symmetry of $\mathbf{J}$ at every state
(Eq.~\ref{eq:skew_sym}) and $\mathbf{R}(\mathbf{x})\succeq0$ everywhere
(Eq.~\ref{eq:r_net}). The fit is supported: a held-out kinematic reconstruction
error of \FitTestMSE\ on three subjects that contributed no training sample,
stable to within $0.3\%$ across seeds (Fig.~\ref{fig:training_robustness}a), with
all five objective terms decreasing to a common loss floor. One dynamical
invariant the twin did not author is reproduced: the stochastic free-run attains
an avalanche branching parameter $\sigma=\LadderSigmaModel$ against the measured
$\LadderSigmaReal$, so near-critical activity is generated by the metriplectic
fluctuation structure rather than fitted.

Two rungs are not cleared, and the failures are quantitative. The aperiodic
spectral exponent of the free-run is $\beta=\LadderOneModel$ against the measured
$\LadderOneReal$ --- a gap of $\Delta\beta\approx0.7$ (Fig.~\ref{fig:spectrum})
--- so the twin's dynamics are spectrally too smooth and carry too little
broadband high-frequency power. The detrended-fluctuation exponent is
$\alpha=\LadderDFAModel$ against the measured $\LadderDFAReal$, above the
stationary band $[0.5,1]$, so the dynamics are over-persistent rather than
itinerant. Rungs~3, 5 and 6 of Table~\ref{tab:validation_ladder} are not scored
at all: no forecasting horizon, no perturbational test of the ports
$\mathbf{G}$, and no cross-session transfer. The three-port controllability
figure of Section~\ref{sec:upgrade} --- about $5\%$ of the $64$-node state space
directly reachable --- bounds what the stimulation interface can currently be
expected to do, independently of any validation.

What the evidence does not support should be stated as plainly. It does not
support the twin as a replica of an individual brain: the fit is across subjects,
without any subject's own anatomical connectome, so it is a population-level
model of the recordings. It does not support the recovered connectome as a
property of cortex rather than of the montage, for the volume-conduction reasons
of Section~\ref{sec:internal_connectome} and Section~\ref{sec:limitations}; the
association between learned coupling magnitude and the empirical alpha prior is
real but weak (Spearman $\rho=0.08$), which is what a suppressive gate should
produce and is not evidence of recovered anatomy. It does not support any claim
about closed-loop neuromodulation in a subject, because the decisive rung is
perturbational and open recordings cannot supply it. And a reader shown the
branching result without the spectral and DFA failures would form a materially
wrong impression, which is why all three appear in
Section~\ref{sec:ladder_results} and in Fig.~\ref{fig:validation_ladder}.

What the framework does supply is a set of ways to be wrong that are specified in
advance: an exponent to be mistimed, an operator property to be violated, a
branching ratio to be missed, a power balance to drift. A model whose output is a
decoding score can be inaccurate; it cannot be refuted. On the present evidence
the honest description is a structure-preserving twin that reproduces one
invariant of cortical dynamics and fails two others in identified ways, with the
mechanisms designed to close those gaps named and testable
(Section~\ref{sec:upgrade}).

\section{Conclusion}

We have presented a \textbf{physics-inspired classical digital twin of BCI data} --- the Cortical GNN-pHNN --- reformulated it for the non-equilibrium regime the recordings exhibit, and fit and scored it on real EEG. 

The framework rests on four elements. First, the twin acts on canonical phasor coordinates $\mathbf{x}=[\bm{\phi},\bm{\omega}]$ extracted from recorded EEG (EEGMMIDB), with the Stuart-Landau normal form of Section~\ref{sec:metriplectic} serving as the motivation for those coordinates rather than as a data generator. Second, a band-stratified graph energy network resolves the storage function into readable frequency sub-energies (Eq.~\ref{eq:energy_decomp}) with a cross-band module that separates direct power coupling from the theta-phase gating of gamma amplitude associated with working-memory indexing. Third, a phase-locking-gated predictor produces the connectome $\mathbf{J}(\mathbf{x})$ with exact skew-symmetry (Eq.~\ref{eq:skew_sym}) at every state, gated by measured alpha-band coherence (Eq.~\ref{eq:plv_gate}) so that coupling is admitted only where phase coherence is present in the data. Fourth, a state-dependent dissipation model (Eq.~\ref{eq:r_net}) captures regionally heterogeneous decay while preserving $\mathbf{R}(\mathbf{x}) \succeq 0$, and an anatomically structured port matrix (Eq.~\ref{eq:G_matrix}) places tDCS, TMS, and DBS at their scalp targets, giving the twin an explicit interface at which closed-loop neuromodulation can be posed.

The substantive advance is the reformulation that lets such a twin be held to the data honestly. Strict passivity is replaced by the non-equilibrium metriplectic steady state (Section~\ref{sec:metriplectic}), in which a reversible and an irreversible generator coexist under exact degeneracy and an explicit metabolic port sustains the resting limit cycle. The dynamics are lifted onto a latent state observed through an explicit EEG emission operator, so the structure governs the modelled cortical state rather than the sensor montage; excitatory and inhibitory compartments give the twin an operating point that a criticality control can tune; a slow arousal compartment supplies the slow timescale; and metastability objectives target the itinerant repertoire the recordings display. Scored free-running against invariants it did not author, the twin reproduces near-critical branching but not the spectral slope or the long-range temporal correlations --- a mixed verdict we report as such, and the upgrades above are aimed squarely at the rungs it fails.

\subsection{Limitations and Future Directions}
\label{sec:limitations}

Three limitations bound the present results and motivate the upgrade programme of Section~\ref{sec:upgrade}. First, the storage function $\mathcal{H}(\mathbf{x})$ is a Lyapunov function, not a directly measurable cortical energy; its absolute value carries no physical units and is not interpretable without calibration against haemodynamic data (fMRI BOLD or fNIRS), for which the metabolic port of the metriplectic reformulation (Eq.~\ref{eq:ness_balance}) is the natural anchor. Second, and most consequentially for what the twin can be said to represent, it is fitted to \emph{scalp} EEG: a low-rank, instantaneously mixed projection of the underlying field. Volume conduction, the common reference, and the lead field are therefore baked into the learned generator, so the recovered connectome is in part a property of the montage rather than of the brain, and the alpha-band PLV gate inherits that inflation (Section~\ref{sec:statespace}).

Third, the validation ladder is only partly cleared: the fitted twin reproduces near-critical avalanche branching ($\sigma\approx1$) but not the $1/f$ spectral slope or the long-range-temporal-correlation (DFA) exponent measured on the recordings (Section~\ref{sec:ladder_results}). Closing that gap requires the excitation--inhibition criticality control and volume-conduction-robust source-space connectivity of Section~\ref{sec:upgrade}, together with a phasor extractor robust to eye-blink and muscle non-stationarities; the perturbational rung (TMS-EEG) remains beyond open data, so the stimulation ports $\mathbf{G}$ are so far untested under an actively delivered perturbation.

Accordingly, Section~\ref{sec:upgrade} lays out the path to a firmer footing, whose priorities are, in order: (i)~adopt the non-equilibrium, metriplectic reformulation (Section~\ref{sec:metriplectic}) and re-verify autonomous limit-cycle sustainment over a long-horizon rollout; (ii)~ground the connectome in volume-conduction-robust connectivity and a diffusion-MRI structural backbone with conduction delays; (iii)~fit and validate on a public dataset (e.g., BCI Competition~IV \cite{Blankertz2007}) against the falsifiable ladder of Table~\ref{tab:validation_ladder}; (iv)~introduce fluctuation--dissipation stochasticity and subject-specific posterior inference; and (v)~establish causal fidelity through TMS-EEG perturbation and closed-loop energy-shaping control (IDA-PBC \cite{Ortega2002}). Modelling of pathological states --- representing seizure onset as a dissipation failure driving hypersynchronous energy growth --- becomes expressible once the metriplectic entropy-production balance (Eq.~\ref{eq:entropy_production}) is in place, since within the twin the anomaly appears as a breakdown of the steady-state balance rather than a mere sign violation.

The Cortical GNN-pHNN thus establishes a rigorous, interpretable, and thermodynamically consistent foundation for a \textbf{classical digital twin} of brain--computer-interface data --- a structure-preserving substrate on which decoding, simulated stimulation, and eventually closed-loop neuromodulation can be designed and stress-tested before they touch a subject. We are deliberate about what it is and is not. The port-Hamiltonian form is a modelling choice that buys passivity, a certified power balance, and a decomposition one can interrogate object by object; it is not an assertion that cortex is a port-Hamiltonian system. And fitted to scalp EEG without an individual's own connectome, the twin is a population-level model of the \emph{recordings}, held to account against invariants it did not author --- not a replica of a particular brain.

\subsection*{Data and Code Availability}

All data used in this study are publicly available. The cortical recordings are
the PhysioNet EEG Motor Movement/Imagery database (EEGMMIDB), collected with the
BCI2000 system \cite{schalk2004} and distributed through PhysioNet
\cite{goldberger2000} ($64$ channels, $12$ subjects, $60$ recordings), retrieved
via the MNE-Python dataset interface. The derived phasor design matrix, the
per-band phase-locking-value priors, and the model-independent dynamical
invariants are computed from these recordings by the pipeline described in
Section~\ref{sec:methods}. The code for phasor extraction, model training, the
validation-ladder scoring, and figure generation is available at
\url{https://github.com/mindverse-computing/Classical-Virtual-Mind}.

\bibliographystyle{unsrt}
\bibliography{references}

\end{document}